\documentclass[useAMS,usenatbib]{mn2e}
\usepackage{graphicx} 
\usepackage{color}

\def\be{\begin{equation}} 
\def\ee{\end{equation}}

\def\gsim{\lower.5ex\hbox{\gtsima}} 
\def\lsim{\lower.5ex\hbox{\ltsima}} \def\gtsima{$\; \buildrel > \over 
\sim \;$} \def\ltsima{$\; \buildrel < \over \sim \;$} \def\prosima{$\; 
\buildrel \propto \over \sim \;$} \def\gsim{\lower.5ex\hbox{\gtsima}} 
\def\lsim{\lower.5ex\hbox{\ltsima}} 
\def\simgt{\lower.5ex\hbox{\gtsima}} 
\def\simlt{\lower.5ex\hbox{\ltsima}} 
\def\simpr{\lower.5ex\hbox{\prosima}} \def\la{\lsim} \def\ga{\gsim} 
  
 \def\gtsima{$\; \buildrel > \over \sim \;$} 
\def\ltsima{$\; \buildrel < \over \sim \;$} 
\def\gsim{\lower.5ex\hbox{\gtsima}} 
\def\lsim{\lower.5ex\hbox{\ltsima}} 
\def\simgt{\lower.5ex\hbox{\gtsima}} 
\def\simlt{\lower.5ex\hbox{\ltsima}} 
\def\simpr{\lower.5ex\hbox{\prosima}} 
\def\la{\lsim} 
\def\ga{\gsim}

\def\E3{{\cal E}_{\rm g}^{III}}

\def\Msun{\rm M_\odot}
\def\Zsun{\rm Z_\odot}

\def\Msun{\rm M_\odot}
\def\myr{\rm Myr}
\def\gyr{\rm Gyr }
\def\Zsun{\rm Z_\odot}
\def\M*{M_*}
\def\Z*{Z_*}
\def\L*{L_*}

\title[Early galaxy formation physics]{Essential physics of early galaxy formation} 
\author[Dayal et al.]{Pratika Dayal$^{1}$\thanks{E-mail:prd@roe.ac.uk (PD)}, Andrea Ferrara$^{2,3}$, James S. Dunlop$^1$ \& Fabio Pacucci$^{2}$ \\ 
$^{{1}}$ SUPA\thanks{Scottish Universities Physics Alliance}, Institute for Astronomy, University of Edinburgh, Royal Observatory, Edinburgh, EH9 3HJ, UK\\
$^{2}$ Scuola Normale Superiore, Piazza dei Cavalieri 7, 56126 Pisa, Italy\\
$^{3}$Kavli IPMU, The University of Tokyo, 5-1-5 Kashiwanoha, Kashiwa 277-8583, Japan}

\begin{document} 
 
\date{} 

\maketitle

\begin{abstract}
We present a theoretical model embedding the essential physics of early galaxy formation ($z \simeq 5-12$) based on the single premise that any galaxy can form stars with  a maximal {\it limiting efficiency that provides enough energy to expel all the remaining gas, quenching further star formation}. This simple idea is implemented into a merger-tree based semi-analytical model that utilises two mass and redshift-independent parameters to capture the key physics of supernova feedback in ejecting gas from low-mass halos, and tracks the resulting impact on the subsequent growth of more massive systems via halo mergers and gas accretion. Our model shows that: \textit{(i)} the smallest halos (halo mass $M_h \leq 10^{10} \Msun$) build up their gas mass by accretion from the intergalactic medium; \textit{(ii)} the bulk of the gas powering star formation in larger halos ($M_h \geq 10^{11.5} \Msun$) is brought in by merging progenitors; \textit{(iii)} the faint-end UV luminosity function slope evolves according to $\alpha = -1.75 \log \,z -0.52$. In addition, \textit{(iv)} the stellar mass-to-light ratio is well fit by the functional form $\log\,  M_* = -0.38 M_{UV} -0.13\, z + 2.4$, which we use to build the evolving stellar mass function to compare to observations. We end with a census of the cosmic stellar mass density (SMD) across galaxies with UV magnitudes over the range $-23 \leq M_{UV} \leq -11$ spanning redshifts $5 < z < 12$: \textit{(v)} while currently detected LBGs contain $\approx 50$\% (10\%) of the total SMD at $z=5$ (8), the {\it JWST} will detect up to 25\% of the SMD at $z \simeq 9.5$.
\end{abstract}


\section{Introduction}
According to the standard cosmological Lambda Cold Dark Matter ($\Lambda$CDM) model 
the early Universe was almost perfectly homogeneous and isotropic, a picture supported by
the small temperature anisotropies measured by the BOOMERanG experiment \citep{lange2001}, 
and the COBE \citep{fixsen1996}, WMAP \citep{hinshaw2013} and PLANCK 
satellites \citep{planck2013}. This (now well-established) model predicts 
that the earliest bound structures were low-mass Dark Matter (DM) halos that formed 
due to gravitational instability in slightly over-dense regions. These low-mass 
structures acted as building blocks and merged to form successively larger structures in a hierarchical sequence.

The mechanical and radiative energy deposition by stars in the earliest galaxies 
affected the subsequent star formation history via a number of physical processes 
collectively referred to as ``feedback''. These include shock-heating and 
ejection of gas out of the DM halos, photo-evaporation and molecule dissociation, 
to mention a few.  In general, these processes suppress and, in some situations, can 
quench further star formation. Feedback is ubiquitously invoked in astrophysics to 
solve problems ranging from galactic to cosmological scales \citep{dubinski-carlberg1991, 
navarro1996, moore1999, klypin1999, moore1999b, springel2008, white-frenk1991, 
yoshida2002, springel-hernquist2003b, gnedin1998, aguirre2001, tornatore2007}.

Given the low DM masses of the earliest galaxies, a tiny amount of star formation 
is sufficient to push out most (or indeed all) of the gas from these systems, 
potentially leading to a complete blow-away \citep{maclow-ferrara1999}. 
These ``feedback-limited" galaxies then have to wait for gas to either be re-accreted 
from the surrounding intergalactic medium (IGM) or to be brought in by mergers to 
re-ignite further star formation. As halos build-up mass with time, their DM 
potential well can sustain much larger star formation rates (SFR) without losing gas. 
This naturally implies that, at any given time, 
there is a \textit{limiting star formation efficiency} such that the energy 
produced by newly-formed stars is sufficient to expel all the remaining gas, 
quenching further star formation (at least temporarily). 

We implement this one simple idea into a semi-analytic model to trace the formation 
and evolution of galaxies over the first billion years of cosmic time, 
from redshift $z=12$ to $z=5$. Our model follows the assembly of galaxies 
through the mergers of their DM progenitors. We trace all major baryonic processes 
including star formation, supernova (SN)-powered gas ejection, gas/stellar mass 
growth through mergers, and gas accretion from the IGM. In the spirit of maintaining 
simplicity to isolate the fundamental physics driving galaxy evolution, our model 
utilises only \textit{two redshift and mass-independent free parameters}: 
(a) the star formation efficiency threshold, $f_*$, and (b) the fraction of SN energy that drives winds, $f_w$. 

The construction of such a simple model is very timely, given the immense amount of data on 
high-$z$ Lyman Break Galaxies (LBGs) that has been acquired over the past few years: 
surveys with the Wide Field Camera 3 (WFC3) on the {\it Hubble Space Telescope} ({\it HST}) 
have revolutionized our understanding of the faintest LBGs, providing unprecedented 
constraints on the faint-end of the evolving ultraviolet luminosity function 
\citep[UV LF; e.g.][]{bouwens2007, mclure2009, bouwens2010a, mclure2010, oesch2010, 
mclure2013} while ground-based wide-field surveys such as UltraVISTA are revealing 
luminous galaxies populating the bright-end of the UV LF at $z \simeq 7$ \citep[e.g.][]{ouchi2010, bowler2012,
bowler2014}. Spectral energy distribution (SED) fitting to 
broad-band photometry has also been used to build galaxy stellar mass functions 
\citep[SMF;][]{gonzalez2011}, understand the stellar populations in these sources through their 
UV ($\beta$) slopes \citep{bouwens2010b, finkelstein2012, wilkins2011, dunlop2012, 
dunlop2013, bouwens2013,rogers2014}, infer their physical properties 
\citep{oesch2010b, labbe2010a, mclure2011, labbe2013} and calculate the growth of stellar 
mass density (SMD) with redshift \citep{labbe2010a, gonzalez2011, stark2013}. 
The Cluster Lensing And Supernova survey with Hubble \citep[CLASH;][]{postman2012} 
is now using galaxies magnified by strong-lensing to further constrain the 
number-density of high-$z$ sources \citep{zheng2012,bradley2013,coe2013} and 
study their nebular emission \citep{smit2014}.

A number of earlier works have attempted to model high-$z$ galaxies using 
cosmological simulations \citep[e.g.][]{finlator2007, dayal2009, nagamine2010, 
dayal2010a, forero2011, salvaterra2011, dayal2012, jaacks2012, dayal2013}. 
Although these studies can shed light on important physical properties of galaxies 
(halo/stellar/gas masses, metallicities), their assembly, and their clustering, 
running simulations necessarily involves a number of assumptions regarding the 
star-formation density threshold, gas ejection, metal pollution and 
fraction of SN energy that can power winds, to name a few. Other models have 
used semi-analytic approaches to reproduce observations invoking multiple 
free-parameters \citep{cole1991, somerville-primack1999, baugh2006, croton2006, delucia2010, benson2012, lu2013}. 

In this work, our aim is instead to build the simplest model, based on only two free parameters than can be readily constrained by existing data
\footnote{The cosmological model used in this work corresponds to the $\Lambda$CDM Universe with DM, 
dark energy and baryonic density parameter values of ($\Omega_{\rm m },\Omega_{\Lambda}, 
\Omega_{\rm b}) = (0.2725,0.702,0.04)$, a Hubble constant $H_0 = 100h= 70 \, {\rm km\, s^{-1} Mpc^{-1}}$, a 
primordial spectral index $n_s=0.96$ and a spectral normalisation $\sigma_8=0.83$, 
consistent with the latest results from the PLANCK collaboration \citep{planck20132}.}, in order to isolate 
the fundamental physics that shapes galaxy evolution in the first billion years. We show that, 
once the two parameters have been fixed, this model naturally yields the 
correct mass-to-light ratios, SMF and SMD, without the need to 
invoke any {\it ad-hoc} free parameters. We make predictions for the fractional contribution 
of galaxies of different luminosities to the SMD, which can be tested in the future with 
the {\it James Webb Space telescope} ({\it JWST}).

\section{Theoretical model}
As introduced above, our aim is to build the simplest model for galaxy formation based 
solely on the balance between the amount of Type II SN (SNII; mass $\geq8 \Msun$) energy available to drive winds, and the 
gravitational potential of the host DM halo. Our idea is as follows: if the SNII kinetic 
energy is larger than the binding energy of a halo, the galaxy will lose all of its gas 
mass and will be unable to form any more stars. However, halos with a binding energy 
larger than the SNII kinetic energy will only lose part of their gas and can 
continue forming stars. This simple idea and its mathematical formulation are detailed in the following. 

\subsection{Feedback-limited star formation efficiency}
\label{fbsec}

Radiative cooling is very efficient in dense low-mass halos at high-$z$. Left unchecked, 
this leads to an over-production of stars and too many baryons being locked up 
in condensed halos (as compared to observations), a problem canonically 
termed  ``overcooling'' \citep{benson2003, springel-hernquist2003b}. This problem 
can be alleviated by introducing SN feedback that reduces the star-formation efficiency of 
small halos by ejecting their gas, quenching further star formation 
\citep[e.g.][]{maclow-ferrara1999, springel-hernquist2003b,greif2007}, as formulated below.

The formation of an amount $M_*(z)$ of stars at redshift $z$ can impart the 
ISM with a total SNII energy $E_{SN}$ given by
\begin{equation}
E_{SN} = f_w E_{51} \nu M_*(z) \equiv f_w v_s^2 M_*(z),
\label{esn}
\end{equation}
where each SNII is assumed to impart an (instantaneous) explosion energy of 
$E_{51}=10^{51}{\rm erg}$ to the ISM and $\nu = [134 \, {\rm \Msun}]^{-1}$ 
is the number of Type II SNe per stellar mass formed for a Salpeter IMF between 
$0.1-100 \Msun$; we use this IMF in all calculations and simply refer to SNII as SN in what follows. 
The values of $E_{51}$ and $\nu$ yield $v_s= 611$ km s$^{-1}$. Finally, $f_w$ is the fraction of the SN explosion energy that is converted into kinetic form and drives winds \footnote{Using the lifetime function of \citet{padovani-matteucci1993}, stars with masses between $9.98 -100 \Msun$ would contribute to the total SN energy for our model time step of 20 Myrs as opposed to the range $8-100 \Msun$ used in this work. This would lead to a value of $\nu = [184 \, {\rm \Msun}]^{-1}$ and $v_s= 521$ km s$^{-1}$. However, this decrease in $v_s$ by a factor of 1.17 does not affect any of our model results, confirming that neglecting the lifetimes of stars that result in SNII is a valid approximation.}. 
 
For any given halo, the energy $E_{ej}$ required to unbind and eject all the ISM gas can be expressed as
\begin{equation}
E_{ej} = \frac{1}{2} [M_{g,i}(z)-M_*(z)] v_e^2,
\label{eej}
\end{equation}
where $M_{g,i}(z)$ is the gas mass in the galaxy at epoch $z$; the term $M_{g,i}(z)-M_*(z)$ 
implies that SN explosions have to eject the part of the initial gas mass not converted into stars. 
Further, the escape velocity $v_e$ can be expressed in terms of the halo rotational velocity,$v_c$, as $v_e = \sqrt 2 v_c$.
 
We then define the {\it ejection efficiency}, $f_*^{ej}$, as the fraction of gas that must be 
converted into stars to ``blow-away" the remaining gas from the galaxy (i.e.  $E_{ej} \le E_{SN} $). This 
can be calculated as 
\begin{equation}
f_*^{ej}(z) = \frac{v_c^2(z)}{v_c^2(z) + f_w v_s^2}.
\label{fej}
\end{equation}
The {\it effective efficiency} can then be expressed as 
\begin{equation}
f_*^{eff} =min[f_*,f_*^{ej}].
\end{equation}
This represents the maximum fraction of gas that can be converted into stars in a galaxy without 
expelling all the remaining gas. Since $v_c$ scales with the halo mass ($M_h$), 
efficient star formers (hosted by large DM halos) can continuously convert a fraction $f_*$ 
of their gas into stars, while feedback-limited systems can form stars with a maximum efficiency 
dictated by $f_*^{ej}$ that decreases with decreasing halo mass. Matching the bright and faint ends of the 
evolving UV LF requires $f_* = 0.03$ and $f_w = 0.1$ as explained in Sec. \ref{sec_phy} below. 

\begin{figure}
\center{\includegraphics[scale=0.48]{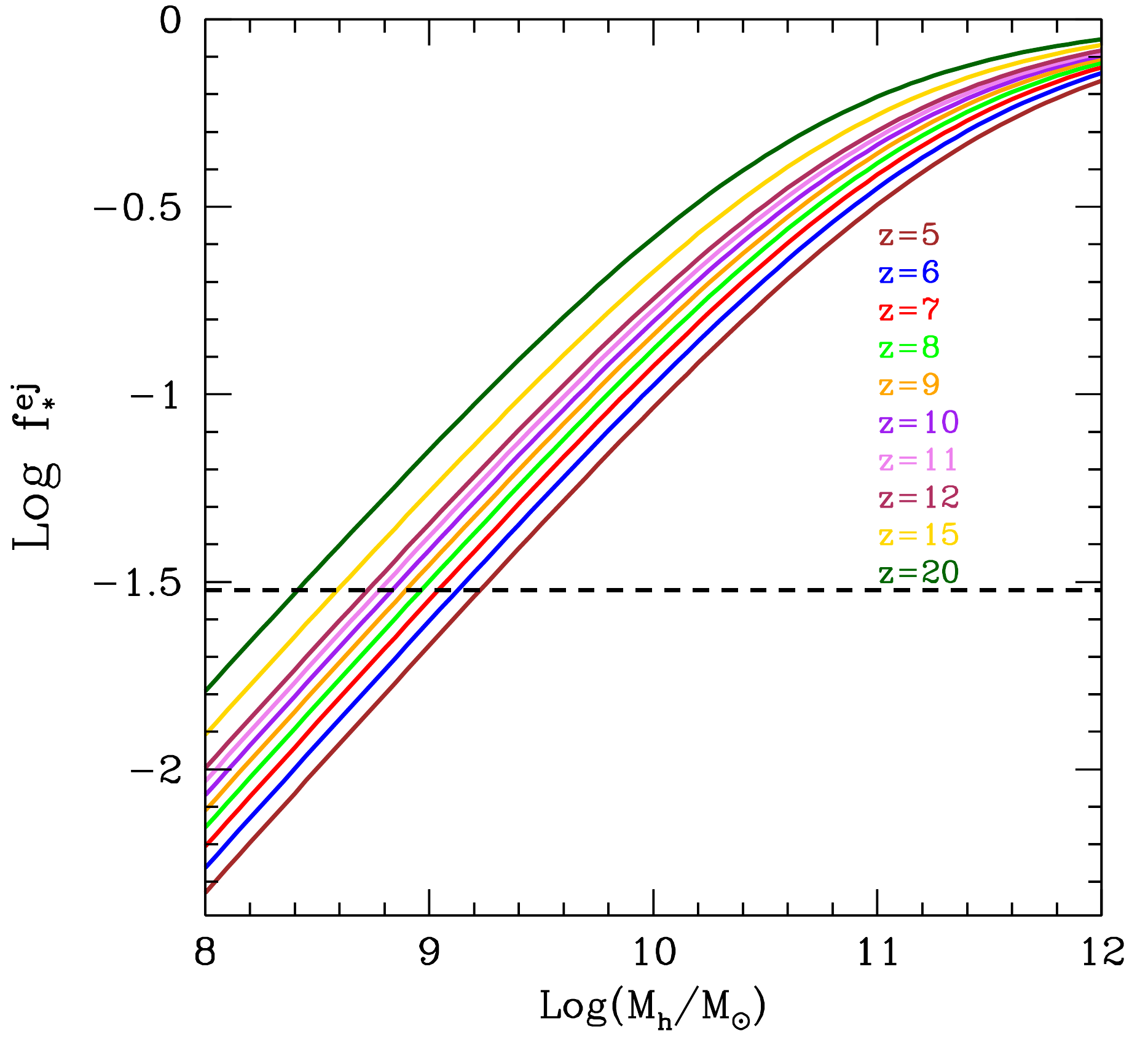}}
\caption{The ejection efficiency ($f_*^{ej}$) as a function of halo mass for $z \simeq 5-20$; 
this is the star-formation efficiency required to eject all the gas from the 
galaxy and quench further star formation. The horizontal line shows $f_* = 0.03$. 
Since $f_*^{eff} =min[f_*,f_*^{ej}]$, galaxies with $f_*^{ej}>0.03$ saturate 
at an effective efficiency of $f_*^{eff} = f_* = 0.03$. We assume each SN imparts an explosion energy of $E_{51}=10^{51}{\rm erg}$, of which a fraction $f_w=0.1$ drives winds and the SN rate ($\nu$) is calculated assuming a Salpeter IMF between $0.1-100\Msun$. These values of $E_{51}$ and $\nu$ lead to $v_s = 611$ km s$^{-1}$ (see Sec. \ref{fbsec} for details).}
\label{feff} 
\end{figure}

Galaxies of a given $M_h$ value are more compact (i.e. have deeper potential wells) and 
rotate faster with increasing redshift as $v_c \propto (1+z)^{1/2}$. Using Eqn. \ref{fej} this 
implies that a given $f_*^{ej}$ value is reached for progressively lower $M_h$ values with 
increasing redshift, as shown in Fig. \ref{feff}. Given that $f_*^{eff} = min[f_*,f_*^{ej}]$, 
this means that $f_*^{eff}$ saturates to $f_*$ for lower $M_h$ values with increasing 
redshift. In other words, galaxies of a given halo mass are more efficient at holding on to 
their gas with increasing redshift as a result of their deeper potential wells. This feedback 
function (behaviour of $f_*^{ej}$ as a function of halo mass) is shown in Fig. \ref{feff} 
for $z=5$ to $z=20$. Quantitatively, while galaxies with masses as low as $M_h \simeq 10^{8.45}\Msun$ 
saturate to $f_* = 0.03$ and become efficient star formers at $z=20$, galaxies 
have to be as massive as $M_h = 10^{9.25}\Msun$ at $z=5$ to achieve the same $f_*$ value.

\subsection{Merger tree physics}
\label{mtsec}
 
\begin{figure*}
\center{\includegraphics[scale=0.5]{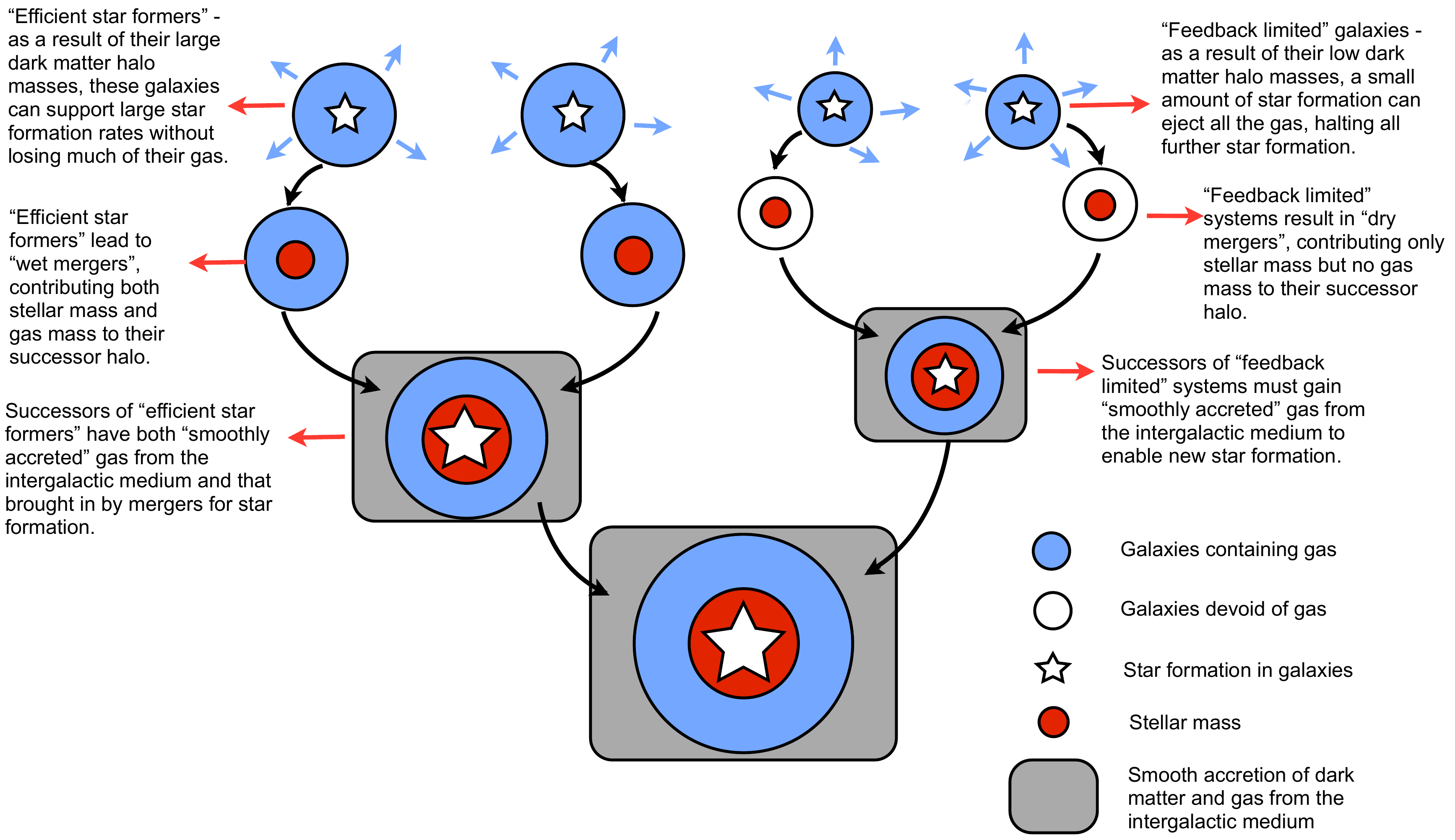}}
\caption{Merger tree showing the assembly of a galaxy in the hierarchical structure formation model. 
Low-mass systems are ``feedback-limited" as SN winds can expel most/all of their gas, quenching 
further star formation. On the other hand, larger-mass systems are ``efficient star-formers" 
and can convert gas into stars at a fixed efficiency ($f_*$) at any given time (see Sec. \ref{fbsec}). 
Galaxies build up in mass hierarchically from the mergers of smaller progenitors which can be 
a combination of feedback-limited systems and efficient star-formers. In addition to 
gaining gas from the mergers of their progenitors, galaxies also ``smoothly accrete" both DM and gas from the IGM. }
\label{mt} 
\end{figure*}

We implement the above simple physical ideas into standard DM halo merger trees tracing 
the formation of increasingly larger systems from the mergers of smaller progenitors as 
shown in Fig. \ref{mt} \citep{white-frenk1991, lacey-silk1991, cole1994}. We build merger 
trees for 800 $z=4$ galaxies equally spaced in $\log\, M_h$ between $10^{8-13}\Msun$ 
using the modified binary merger tree algorithm with accretion presented in 
\citet{parkinson2008}. In brief, the merger tree for each simulated DM halo 
starts at $z=4$ and runs backward in time up to $z=20$, with each halo fragmenting 
into its progenitors. At any given time-step, a halo of mass $M_0$ can either lose 
a part of its mass (i.e. fragment into halos below the mass resolution limit $M_{res}$) or 
fragment into two halos with masses $M_{res} < M < M_0/2$. The mass below the resolution 
limit then accounts for ``smooth-accretion" from the IGM, in which the halo is embedded. 
We run our merger tree using 70 steps equally spaced in time (by 20 Myrs) and with a 
resolution mass $M_{res} =10^8 \Msun$. Each of the simulated $z=4$ halos is associated 
with the correct number density by matching its halo mass to the Sheth-Tormen 
mass function \citep{sheth-tormen1999}. Then, at any redshift, 
every progenitor is given the same number density as its $z=4$ successor. The use of the conditional mass function given by the extended Press-Schechter theory \citep{bower1991, lacey-cole1993} and the modifications introduced by \citet{parkinson2008} ensure that progenitor halo mass function matches the Sheth-Tormen mass function at any $z$ \citep[see Sec. 2.1][]{dayal2014b}.

Once the merger tree for each galaxy has been constructed, we implement our baryonic physics 
model within it. Given that the baryonic properties of parent halos depend on those of their 
progenitors at earlier times, we now proceed forward in time from $z=20$ and follow the joint 
halo/galaxy evolution. We start from the first DM progenitor (with halo mass $M_0$) along a 
branch of the merger tree and assume that it has an initial gas mass $M_{g,i} (z) = (\Omega_b/\Omega_m) M_0 (z)$. 
A fraction of this gas mass gets converted into a (newly formed) stellar mass $M_*(z)$, such that
\begin{equation}
M_*(z) = f_*^{eff} M_{g,i}(z).
\label{mstarnew}
\end{equation}
In the spirit of maintaining simplicity, we assume that every stellar population has a 
fixed metallicity of $0.05 \Zsun$ and each newly-formed stellar population has an age 
of $2 \, \myr$. Using these parameters with the population synthesis code 
{\small STARBURST99} \citep{leitherer1999}, the UV luminosity at $\lambda = 1500$\,\AA\ 
produced by a newly-formed stellar mass can be expressed as
\begin{equation}
L_{UV} = 10^{33.077} \bigg(\frac{M_{*}}{\Msun} \bigg) \,\, {\rm erg\, s}^{-1} {\rm \AA}^{-1}.
\label{lumnew}
\end{equation}
This star-formation episode must then result in a certain amount of gas $M_{g,ej}(z)$ 
being ejected from the galaxy at the given $z$-step. The fraction of gas mass ejected depends on the ratio of the SN kinetic energy available ($E_{SN}$) and the potential energy required to unbind the gas not turned into stars ($E_{ej}$) which can be expressed as (see Eqns. \ref{esn} and \ref{eej})
\begin{equation}
\frac{E_{SN}}{E_{ej}} = \frac{f_w v_s^2 f_*^{eff} M_{g,i}(z)}{M_{g,i}(z) (1-f_*^{ej}) v_c^2} = \frac{f_w v_s^2 f_*^{eff}}{(1-f_*^{ej}) v_c^2}.
\end{equation}
Substituting $f_*^{ej}$ from Eqn. \ref{fej}, we obtain 
\begin{equation}
\frac{E_{SN}}{E_{ej}} = \frac{f_*^{eff}}{f_*^{ej}}.
\label{raten}
\end{equation}
The value of $M_{g,ej}(z)$ therefore depends on 
whether $f_*^{eff} = f_*$ or $f_*^{eff} = f_*^{ej}$: in the former case the galaxy is 
an ``efficient star-former" that can support a large amount of stellar mass being formed 
without losing much of its gas, while it is a ``feedback-limited" system in the latter 
case with all of its ISM gas being blown-away (see also Fig. \ref{mt}). Using Eqn. \ref{raten}, $M_{g,ej}$ can be mathematically expressed as
\begin{equation}
M_{g,ej}(z) = [M_{g,i}(z) - M_{*}(z)] \frac{f_*^{eff}}{f_*^{ej}},
\label{mej}
\end{equation}
since the initial gas mass is reduced by the amount that is converted into stars. 
The final gas mass, $M_{g,f}(z)$, remaining in the galaxy at that time-step can then be expressed as
\begin{equation}
M_{g,f}(z) = [M_{g,i}(z) - M_*(z)] \bigg[1-\frac{f_*^{eff}}{f_*^{ej}}\bigg].
\label{mf}
\end{equation}
 
On the other hand, a galaxy inherits a certain amount of stars and gas from its progenitors 
following merging events. In addition, this galaxy also obtains a part of its DM (and gas) 
mass through ``smooth-accretion" from the IGM. Consider, for example, a galaxy of halo mass $M_0$ at redshift $z$ that 
has progenitors with halo masses $M_1$ and $M_2$ at redshift $z + \Delta z$. 
The difference between the sum of the progenitor masses and $M_0$ then yields 
the unresolved halo mass that is smoothly-accreted from the IGM, such that $M_{h,acc}(z) = M_0 - (M_1+M_2)$. 
We then make the simple (and reasonable) assumption that the smoothly-accreted DM pulls in a cosmological 
ratio of gas mass with it such that the accreted gas mass is $M_{g,acc}(z) = (\Omega_b/\Omega_m) M_{h,acc}(z)$. 
Thus, the total initial gas mass in the galaxy at $z$ is the sum of the 
newly accreted gas mass, as well as that brought in by its merging progenitors, i.e.
\begin{equation}
M_{g,i}(z) = M_{g,acc}(z) + \sum M_{g,f}(z+\Delta z).
\end{equation}
This $M_{g,i}(z)$ value is then used to calculate the new stellar mass formed in the galaxy 
as described by Eqn. \ref{mstarnew}. The total stellar mass in this galaxy is now the sum 
of mass of the newly-formed stars, and that brought in by its progenitors such that
\begin{equation}
M_{*,tot}(z) = M_*(z) + \sum M_*(z+\Delta z). 
\end{equation}
Eqs. \ref{mej} and \ref{mf} are again used to obtain the ejected, and final gas masses at the given $z$-step. 

Finally, the total UV luminosity of the galaxy is a sum of the new 
luminosity as well as that brought by its progenitors,
\begin{equation}
L_{UV,tot} = L_{UV}(z) +  \sum L_{UV,*}(z+\Delta z).
\end{equation}
Using {\small STARBURST99}, we find that the UV luminosity for a burst of stars 
(normalized to a mass of $1 \Msun$ and metallicity $0.05\, \Zsun$) decreases with time as 
\begin{equation}
\log \bigg(\frac{L_{UV}(t)}{{\rm erg\, s}^{-1} {\rm \AA}^{-1}} \bigg) = 33.0771 - 1.33 \log (t/t_0)  + 0.462,
\label{luv}
\end{equation}
where $t$ is the age of the stellar population (in yr) at $z$ and $\log (t_0/{\rm yr}) = 6.301$; 
we remind the reader that by construction, each newly-formed stellar population has an age of $2 \, \myr$. 

These simple ideas of halo mass/stellar growth, star formation and its associated feedback 
and gas mass ejection/accretion/merged are implemented into our merger tree, 
tracing the growth of galaxies from $z=20$ to $z=4$. As shown in Fig. \ref{mt}, in this model 
low-mass galaxies are {\it feedback-limited star formers}: 
star formation with a low efficiency is sufficient to eject all the gas from these systems, 
quenching further star formation. These feedback-limited galaxies then have to wait for gas 
to either be accreted from their surrounding IGM or for gas to be brought in by mergers to 
re-ignite further star formation. On the other hand, massive galaxies are efficient star formers 
that can sustain much larger star formation rates at a fixed efficiency $f_*$. As expected, 
both low and high-mass galaxies assemble from building blocks that are a combination of 
``feedback-limited" and ``efficient" systems, with galaxies of progressively lower 
masses becoming efficient star formers with increasing redshift, as explained in Sec. \ref{fbsec}.

As seen from the equations in this Section, our fiducial model only has two free 
parameters - the threshold SF efficiency ($f_*$) and the fraction of SN energy 
that drives winds ($f_w$), both of which are {\it independent of redshift and halo mass}.

\section{UV luminosity function}

Once we have implemented the baryonic physics described in Sec. \ref{mtsec} into 
the merger tree and inferred the UV luminosity of each galaxy, we can construct the 
evolving LBG UV LF to be compared with observations, and shed light on the physics that shapes it.

\begin{figure*} 
\center{\includegraphics[scale=0.95]{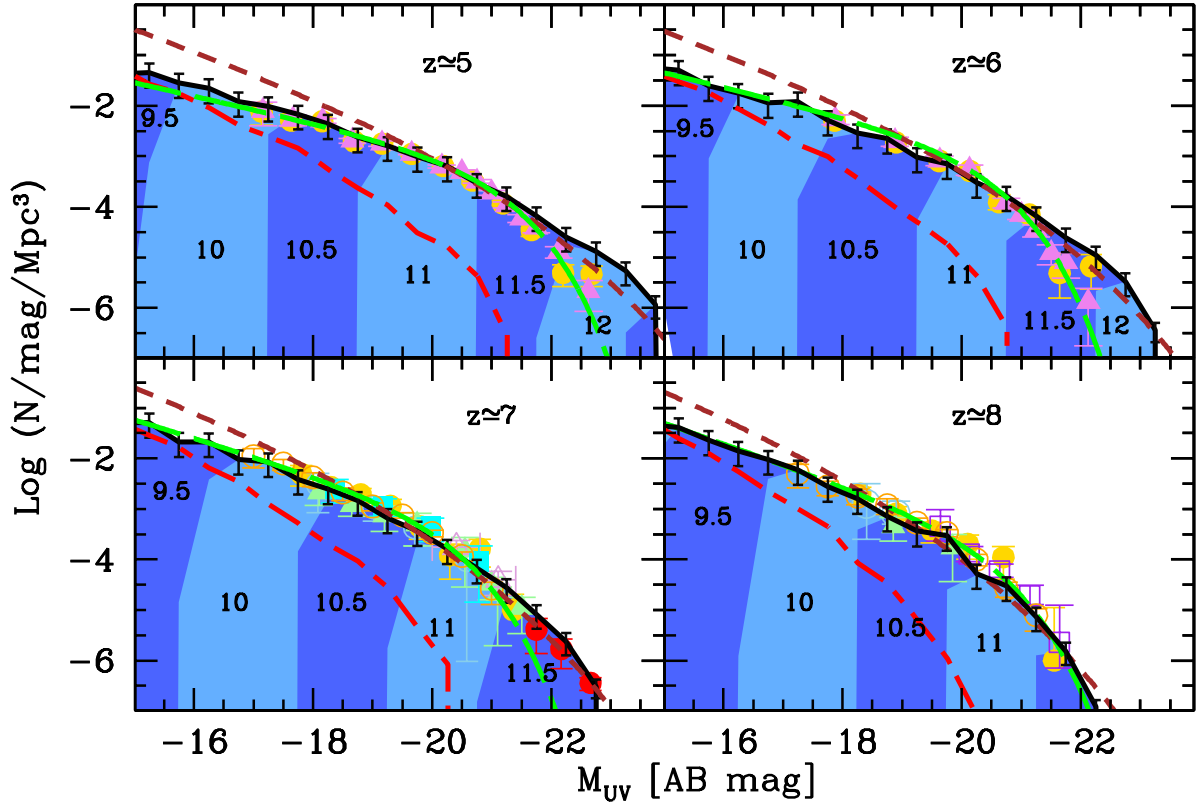}}
\caption{The evolving LBG UV LF at $z \simeq 5-8$. In all panels, the black solid line shows 
the results using our fiducial model, i.e. including gas accretion from progenitors 
and the IGM, and SN-powered gas ejection; the black error bars show poissonian errors 
arising from the luminosity dispersion in each bin. The dashed red line shows that the 
UV LF would have been severely under-estimated (with increasing brightness) 
had we not considered the gas that is brought in by mergers. The brown line shows 
the results obtained by multiplying the halo mass function with a constant star-formation 
efficiency appropriate for the redshift considered (column 4 of Table \ref{table1}). 
In all panels, the dashed green line shows the observationally inferred best-fit 
Schechter function \citep {mclure2009, mclure2013} and points show observational 
results: (a) $z\simeq 5$: \citet[filled circles]{bouwens2007} and 
\citet[filled triangles]{mclure2009}; (b) $z\simeq 6$: \citet[filled circles]{bouwens2007} 
and \citet[filled triangles]{mclure2009}; (c) $z \simeq 7$: \citet[filled squares]{oesch2010}, 
\citet[empty blue circles]{bouwens2010a}, \citet[filled circles]{bouwens2011b}, 
\citet[empty triangles]{castellano2010}, \citet[filled triangles]{mclure2010}, 
\citet[empty orange circles]{mclure2013} and \citet[filled red circles]{bowler2014};  
(d) $z \simeq 8$: \citet[empty blue circles]{bouwens2010a}, \citet[filled circles]{bouwens2011b}, 
\citet[filled triangles]{mclure2010}, \citet[empty squares]{bradley2012} and 
\citet[empty red circles]{mclure2013}. The numbers in the shaded areas under the UV LF 
show the central value of the DM halo mass bin hosting the galaxy as obtained from the fiducial model.}
\label{fig_lbg_uvlf} 
\end{figure*}

\subsection{Understanding the UV LF evolution}
\label{sec_phy}
We start by discussing the behaviour of the UV LF inferred from the halo mass function (HMF). 
We can attempt to construct a simple (and likely unphysical) model of the UV LF 
by multiplying the HMF with a star-formation 
efficiency value $\eta$ (column 4 of Table 1), chosen to match to the knee of the observed UV LF. While $\eta$ is redshift-dependent, its value is constant (i.e. independent of halo mass) at any given redshift. 
This is equivalent to a model without feedback wherein every galaxy contains a cosmological 
ratio of baryons to DM equal to $\Omega_b/\Omega_m$, leading to the observed UV LF tracing 
the same shape as the HMF. Quantitatively, it transpires from the required efficiency 
scaling with redshift, that galaxies of the same UV luminosity reside in halos that are twice as 
massive at $z \simeq 5$ as at $z \simeq 8$ leading to a steepening in the predicted faint-end slope ($\alpha_{HMF} = -2.11 \rightarrow -2.32$ from $z=5 \rightarrow 8$). This simple comparison serves to emphasize the important 
role of feedback; a viable model requires that DM halos form stars with an efficiency that progressively decreases with halo mass

Given that galaxies of a given luminosity are hosted in lower-mass halos at earlier times, one might then expect them to enter the feedback-limited regime producing a flattening of the LF faint-end towards high-$z$ compared to the underlying evolving HMF. It is therefore intriguing to see that the faint-end slope of the theoretical UV LF also steepens with increasing redshift as shown in column 5 of Table \ref{table1}; within errors the faint end slope of the theoretical UV LF remains constant from $z \simeq 8$ to $z \simeq 5$ with respect to the underlying HMF, as shown in Fig. \ref{fig_lbg_uvlf}.

\begin{figure*} 
\center{\includegraphics[scale=0.95]{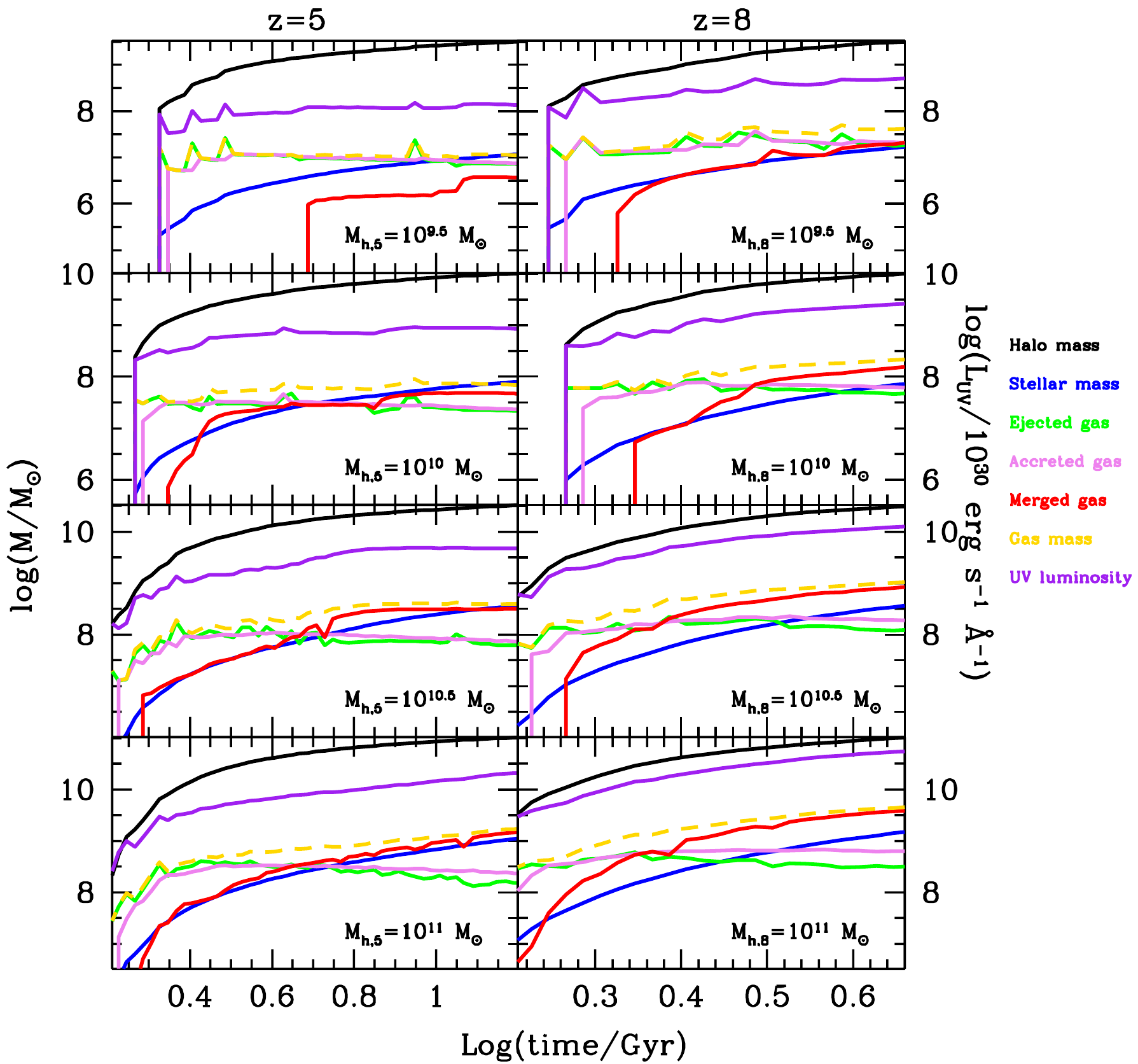}}  
\caption{Galaxy assembly as a function of cosmic time. For the halo mass 
value given in each panel, we show the assembly of galaxies with halo masses 
in the range $10^{9.5}-10^{11} \Msun$ at $z=5$ (left column) and $z=8$ (right column). 
Each panel shows the total halo and stellar mass (in $\Msun$) built-up by a certain 
time; the ejected/accreted/merged and final gas mass values (in $\Msun$) are 
instantaneous quantities. We also show the instantaneous UV luminosity 
in units of $\log(L_{UV}/{\rm erg \, s^{-1} \AA^{-1}}$) which can 
be treated as a proxy for the SFR; the UV luminosity has been uniformly 
scaled down by $10^{30}$ for display reasons. Note that the time-axis is different 
in the left and right-columns: the Universe is twice as old at $z=5$ with an age of $1.21\, 
\gyr$ as compared to $z=8$ where the age is $0.66 \, \gyr$.  }
\label{assembly} 
\end{figure*}

The physical reason for the constant faint-end slope difference between the UV LF and the HMF can be explained as follows. As seen from the shaded regions in Fig. \ref{fig_lbg_uvlf}, the host halo masses increase by a factor $\approx 3$ from $M_h = 10^{9.5} \, (10^{11}) \Msun$ for $M_{UV} = -16 \, (-21)$ at $z=8$ to $10^{10} \, (10^{11.5}) \Msun$ for $M_{UV} = -16 \, (-21)$ at $z=5$. However, a galaxy of a given halo mass has half the time to assemble by $z =8$ when the age of the Universe is about $0.66 \, \gyr$ compared to the $1.21 \, \gyr$ available by $z=5$. As a result of the shorter time available, galaxies of a given halo mass assemble at twice the rate leading to their SFRs (and luminosities) being twice as large at $z=8$ compared to $z=5$. So, although the HMF shifts to progressively lower masses with increasing redshift, the fact that galaxies of a given mass are more efficient at forming stars with increasing redshift helps maintain a constant slope offset between the UV LF and underlying HMF, as shown in Fig. \ref{fig_lbg_uvlf}. 

In summary, our model reproduces both the slope and amplitude of the UV LFs 
as measured by a number of surveys remarkably well, as 
shown in Fig. \ref{fig_lbg_uvlf} and quantified in Table \ref{table1}. 
The strength of our model lies in the fact that it yields UV LFs over $9.5$ ($7$) UV magnitudes at $z=5$ (8) 
using only two time- and mass-independent free parameters $f_*=0.03$, $f_w=0.1$. 
These two parameters together shape the UV LF: while $f_w$ affects the faint-end 
slope of the UV LF where feedback is most effective, $f_*$ determines the amplitude 
and normalization at the bright-end where galaxies can form stars at the 
maximum allowed $f_*$ value. Of course, since $f_*^{eff} =min[f_*,f_*^{ej}]$, 
its value depends both on the halo mass and redshift through the 
dependence of $f_*^{ej}$ on these two quantities as explained in Sec. \ref{fbsec}.  

\begin{table} 
\begin{center} 
\caption {For the redshift shown in column 1, we show the observed faint-end 
slope of the UV LF \citep{mclure2009, mclure2013} in column 2. Columns 3 shows 
the faint-end slope of the UV LF obtained by scaling the HMF using the star-formation 
efficiency shown in column 4. Column 5 shows the faint-end UV LF slope values 
obtained from our fiducial model. The faint-end slopes for the scaled HMF and 
theoretical UV LF have been computed over the absolute 
magnitude range $-18 \leq M_{UV} \leq -12$.} 
\begin{tabular}{|c|c|c|c|c}
\hline 
$z$& $\alpha_{obs}$ & $\alpha_{HMF}$ & $\eta$ & $\alpha_{th}$ \\  
\hline 
$5$& $-1.66^{+0.66}_{-0.66}$ & $-2.11 \pm 0.07$ & $0.006\phantom{0}$ & $-1.74 \pm 0.48$  \\ 
$6$& $-1.71^{+0.11}_{-0.11}$  & $-2.19\pm 0.10$ & $0.0075$ & $-1.89 \pm 0.71$\\ 
$7$& $-1.90^{+0.14}_{-0.15}$ & $-2.25 \pm 0.13$ & $0.009\phantom{0}$ & $-2.02 \pm 0.74$\\ 
$8$& $-2.02^{+0.22}_{-0.23}$ & $-2.32 \pm 0.15$ & $0.011\phantom{0}$ & $-2.10 \pm 0.62$  \\ 
$9$& $-$   & $-$ & $-$ & $-2.19 \pm 0.23$\\ 
$10$& $-$  & $-$ & $-$ & $-2.25 \pm 0.48$ \\ 
$11$& $-$  & $-$ & $-$ & $-2.32 \pm 0.34$\\ 
$12$& $-$ & $-$ & $-$ & $-2.61\pm 0.83$ \\ 
\hline
\label{table1} 
\end{tabular} 
\end{center}
\end{table}

Further, our model naturally predicts that the bright-end of the UV LF at $z \simeq 7$ 
must be flatter than the usually-assumed Schechter function and is compatible 
with both the HMF and the double power-law (DPL) slope estimated 
from the widest-area survey carried out so far at this redshift \citep{bowler2014}. 
Our model slightly over-predicts the number of bright galaxies at $z \simeq 5$ and $z \simeq 6$, 
as can be seen from the same figure. Whether this discrepancy is due to physical 
effects that have been ignored (e.g. dust attenuating the luminosity from 
these massive galaxies \citep{dayal2009}, halo mass quenching \citep{peng2010} 
and/or AGN feedback), or is in fact due to remaining issues with the data analysis 
(e.g. the application of inadequate aperture corrections 
when undertaking photometry of the brightest high-redshift galaxies)
remains a matter for further study. By virtue of reproducing the observed UV LF, the redshift evolution of the star formation rate density (SFRD) predicted by our model is in excellent agreement with high-$z$ SFRD observations \citep{ellis2013}.

Finally, as shown in \citet{dayal2013}, 
we clarify that the evolution of the UV LF is a {\it combination} of luminosity 
and density evolution that depends on the luminosity range probed: 
the evolution at the bright end is genuine luminosity evolution, driven by the brightest 
galaxies continuing to brighten further with time; 
the evolution at the faint end is a mix of positive and negative 
luminosity and density evolution as these tiny systems brighten 
and fade in luminosity, and continually form and merge into larger systems. 

\subsubsection{Faint end galaxies: starving or inefficient?}

A natural question that arises at this point is whether the faint-end 
of the UV LF lies below that which would be inferred from the HMF because 
the fainter galaxies are fuel-supply limited (``starving'') as a result of 
their progenitors having ejected most/all of their gas content, or because they themselves are 
star-forming efficiency-limited (i.e. $f_*^{eff}<f_*$) due to their low masses. This question can easily be answered using Fig. \ref{feff}: 
as shown there, galaxies with $M_h \geq 10^{9}\,, (10^{9.25}) \Msun$  at $z=8 \, (5)$ 
can form stars at the maximum allowed efficiency of $f_*^{eff}= f_* = 3\%$. 
Given that, from our fiducial model, galaxies brighter than $M_{UV}=-15$ are 
hosted in halos more massive than $10^{9.5} \Msun$ at all $z=5$ to $z=8$ 
(Fig. \ref{fig_lbg_uvlf}), galaxies {\it on the currently observable UV LF 
are not efficiency limited}; their luminosities are depressed as a result of 
their progenitors having lost most/all of their gas content, resulting 
in a gas mass that is lower than the cosmological baryon fraction, i.e. starvation. 
The observed UV LF thus holds imprints of the entire past gas build-up history of its progenitors.

\subsubsection{Gas supply: mergers or accretion?}

This bring us to the question of how these galaxies build up their 
gas and stellar content: is their assembly dominated by the gas brought 
in by mergers or that accreted from the IGM? A first hint can be obtained 
from Fig. \ref{fig_lbg_uvlf} where, in the dashed red lines, we show 
the theoretical UV LF that results from assuming each galaxy loses all 
its gas content at every $z-$step resulting in dry mergers (although 
we use the same $f_*^{eff}$ values as the fiducial model). As can be seen, 
while similar to the fiducial model at the faint end, this UV LF drops off 
very steeply with increasing luminosity. This implies that gas brought 
in by mergers is not very important at the faint end since the tiny 
progenitors of these galaxies are feedback limited even in the 
fiducial model, resulting in largely dry mergers. However, the 
progenitors of increasingly luminous galaxies are not feedback limited, 
and contribute to building up a large gas reservoir that powers star 
formation. Indeed, the LF at the bright end drops to 10\% of the fiducial 
model value ($M_{UV}$ decreases by $\approx 2$ magnitudes) if the gas 
brought in by mergers is not taken into account.

To further elucidate the above point, we now examine in detail four 
representative galaxies with halo masses $M_h = 10^{9.5,10,10.5,11}\Msun$ 
at two epochs corresponding to $z=5$ and $z=8$ (see Fig. \ref{assembly}). 
As expected from the hierarchical model, the earlier a galaxy starts building 
up, the larger its final mass can become \citep{dayal2013}, i.e. the progenitors 
of the most massive galaxies start assembling earliest with the progenitors of 
progressively less massive galaxies forming later. As shown in Fig. \ref{assembly}, 
while the progenitors of $M_h = 10^{11} \Msun$ galaxies start forming about 
$0.2 \, \gyr$ after the Big Bang, the progenitors of $10^{9.5} \Msun$ 
galaxies appear only $\approx 0.2 \,\gyr$ later than that. 
Moreover, the progenitors of galaxies as small as $10^{9.5}\Msun$ are 
feedback-limited and bring in negligible amounts ($\simlt 10$\% of the gas acquired 
by accretion) of gas when the galaxy starts assembling. As a result, the 
initial stellar mass build up is dominated by self-accreted gas, 
with mergers becoming important only at the later stages. Gas 
brought in by mergers starts dominating progressively earlier for more 
massive halos; while episodes of star formation are essentially driven by 
minor mergers, massive galaxies build up their gas reservoir from both 
accretion and mergers. Finally, it is interesting to note that galaxies of 
similar halo mass produce a similar amount of stars at both $z=5$ and $z = 8$. 
As galaxies at $z=8$ have about half the time to assemble as compared to 
galaxies at $z=5$, it follows that the specific star formation rate 
correspondingly decreases with time. 

To summarize, the {\it faint-end of the LF is feedback-suppressed} below the 
HMF since star formation in the tiny progenitors of these galaxies led to 
complete gas ejection, reducing their gas mass below the cosmological baryon 
fraction. Smooth accretion from the IGM dominates over mergers in assembling 
the gas mass of the faintest galaxies on the UV LF. Mergers (smooth-accretion) become 
progressively more (less) important with increasing luminosity, with 
{\it mergers supplying most of the gas mass for galaxies at the bright end} of the LF. 
Given their halo masses, while $z=5-8$ galaxies brighter than $M_{UV}=-15$ can 
form stars at the maximum allowed efficiency, their lower luminosities arise 
as a result of their low baryon fraction (since their progenitors are feedback-limited), 
i.e. these galaxies are {\it not star-formation efficiency limited, but starved due to limited fuel supply}.

\subsection{Faint-end slope evolution}
\label{uv9212}

\begin{figure}
\center{\includegraphics[scale=0.48]{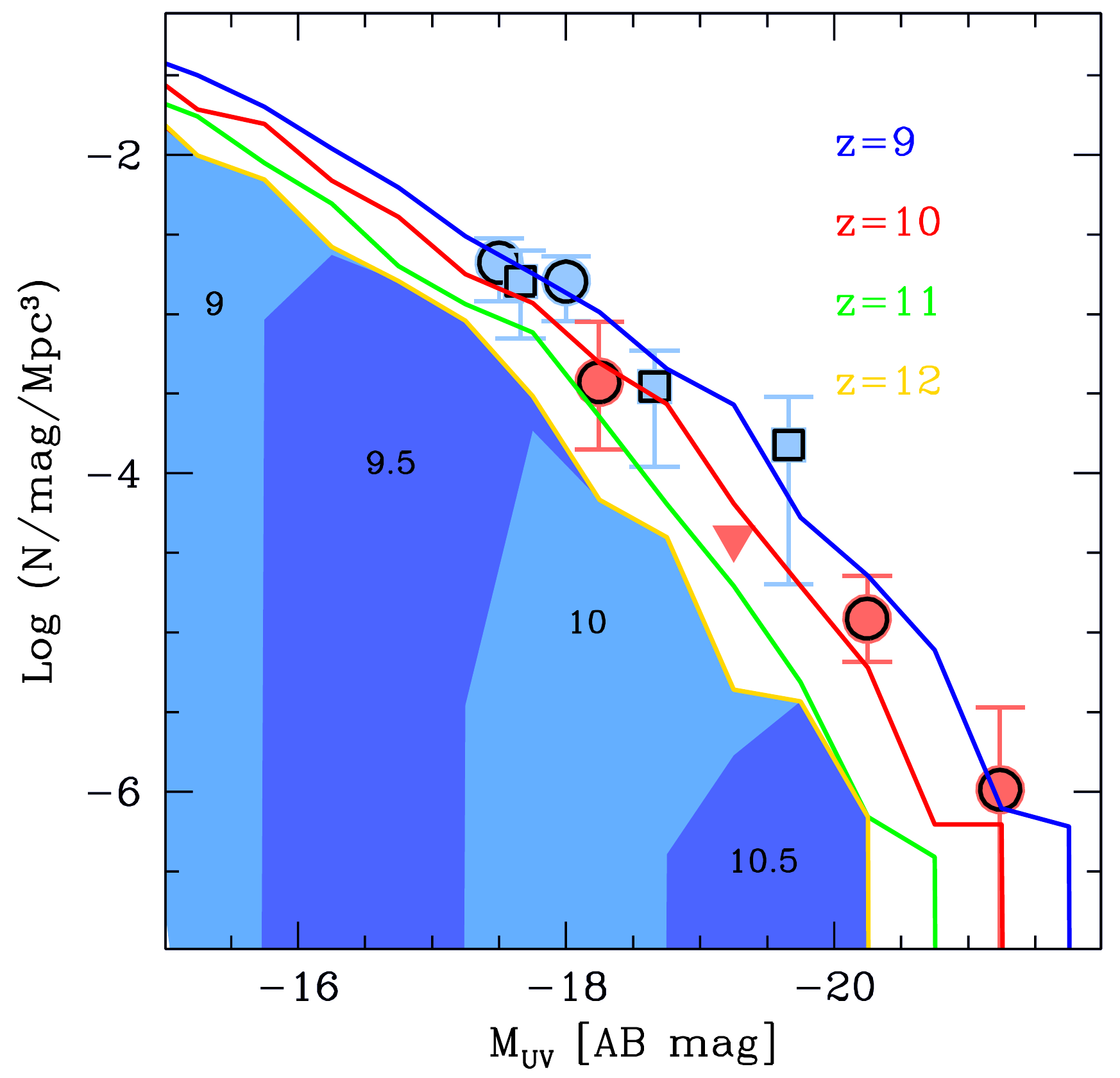}}
\caption{The evolving LBG UV LF at $z \simeq 9-12$: lines show our model results 
with data points showing observational results. The UV LF from our fiducial model is 
in excellent agreement with observations at $z \simeq 9$: \citep[filled blue circles]
[]{mclure2013} and \citep[filled blue squares][]{oesch2013} and at $z \simeq 10$: 
\citep[filled red circles][]{bouwens2014}; the downward pointing triangle represents 
the upper-limit of the $z\simeq 10$ data at $M_{UV} \simeq -19.25$. We present 
our predictions for $z =11,12$ that should be testable with the {\it JWST}. 
The numbers in the shaded area under the $z=12$ UV LF show the mass of the 
DM halo hosting the galaxy. With $M_h \geq 10^9 \Msun$, galaxies as 
faint as $M_{UV} =-15$ at $z=12$ are not star-formation efficiency limited; see Sec. \ref{uv9212}.}
\label{uv9} 
\end{figure}

Buoyed by the success of our model in reproducing the observed UV LFs 
from $z=5$ to $8$, we extend our results to redshifts as high as $z=9 - 12$ (Fig. \ref{uv9}), 
and provide a functional form for the redshift-evolution of the faint-end 
slope, $\alpha$ (also see Table  \ref{table1}).

It is encouraging to see that the UV LF predicted by our fiducial model 
is in agreement with the sparsely-sampled UV LF at $z=9$ \citep{mclure2013,oesch2013}. 
We remind the reader that we use the same values of our two free parameters 
at all $z$ such that $f_* = 0.03$ and $f_w = 0.1$, i.e. we do not invoke any {\it redshift-dependent ad-hoc free parameters}. Further, the $z=10$ theoretical 
UV LF is also consistent with the UV LF inferred by \citet{bouwens2014}. 
However, our predictions are slightly higher than the observational upper limits 
inferred by these authors at $M_{UV} =-19.5$. Improved data are needed at this epoch 
to determine whether this apparent drop in the UV LF is real.

Galaxies at $z=9-12$ more luminous than $M_{UV}=-15$ are hosted in halos 
with mass $M_h \geq10^{9} \Msun$ as shown by the shaded regions 
(for $z=12$) in Fig. \ref{uv9}. From Fig. \ref{feff}, we see that 
galaxies with masses as low as $M_h  = 10^{8.75} \Msun$ start forming 
stars with the maximum allowed efficiency of $f_* = 0.03$. This implies 
that the galaxies on the UV LF at $z=9-12$ are not themselves efficiency-limited 
but fuel-supply limited; the progenitors of galaxies at the faint-end 
were feedback limited, reducing their gas mass below the cosmological baryon fraction.

A by-product of our model is the prediction of the redshift-evolution of the 
faint-end slope of the UV LF. The exact value of this slope and its redshift-evolution 
is important, given that it is the galaxies occupying the faint-end of the 
UV LF that provide most of the photons for cosmic reionization 
\citep[e.g.][]{choudhury-ferrara2007, salvaterra2011}. 
As explained in Sec. \ref{sec_phy}, the UV LF steepens with increasing 
redshift from $z=5$ to $12$, mirroring the behaviour of the scaled HMF 
which shifts to progressively lower masses. From our fiducial model 
we provide a simple functional form for our predicted redshift evolution of 
$\alpha$ (which can be used in reionization calculations):
\begin{equation}
\alpha = -1.75 \, \log\, z - 0.52.
\end{equation}
This relation is valid over two orders of magnitude in luminosity,
for the magnitude range $-18 \leq M_{UV} \leq -12$ and over the redshift range 
$z=5$ to $z = 12$.

\begin{figure*} 
\center{\includegraphics[scale=0.85]{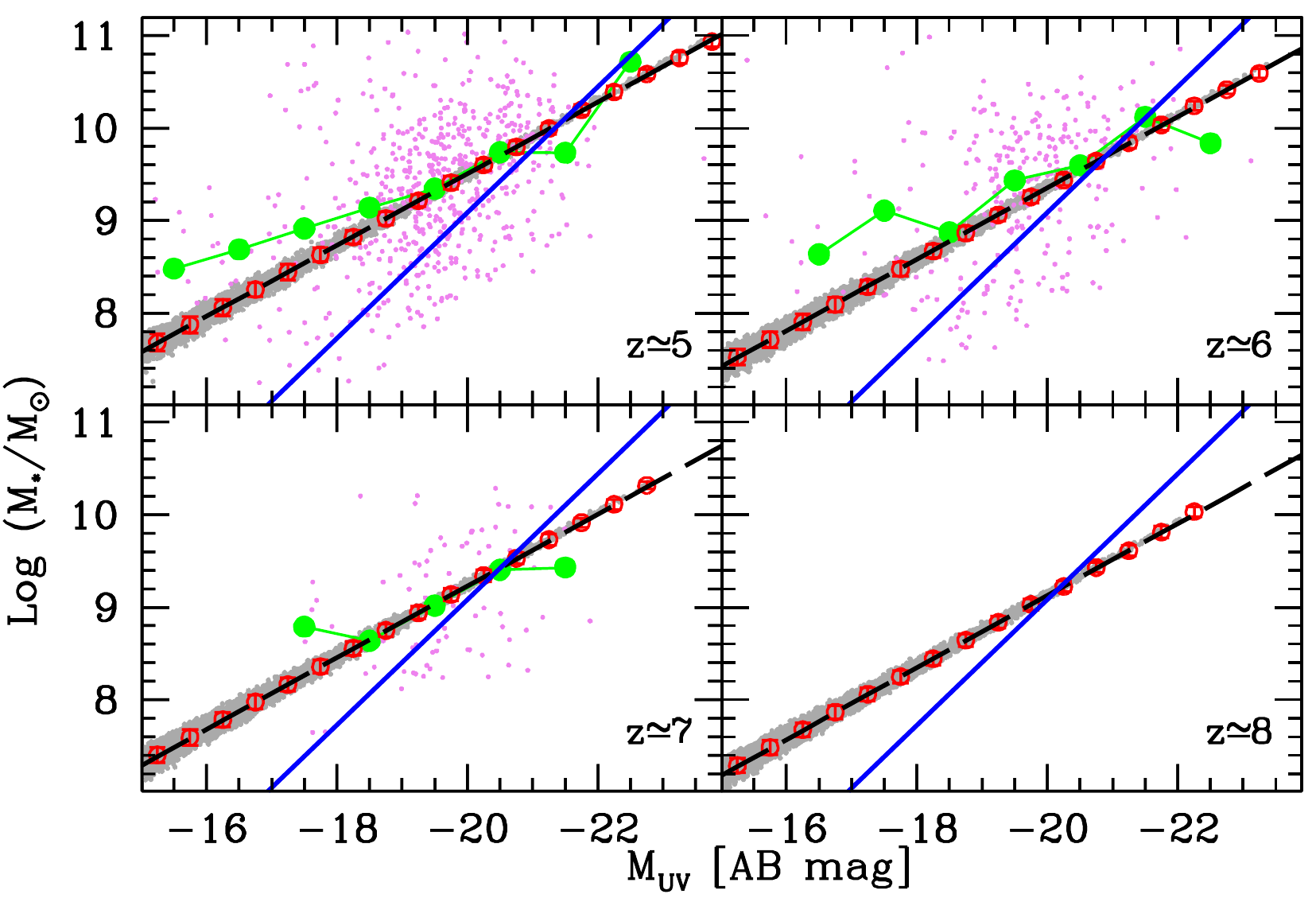}} 
\caption{Mass-to-light relation showing galaxy stellar mass as a 
function of UV magnitude. Red points show the predicted average $M_*$ value 
in each UV bin together with the $1\sigma$ error, and gray points show 
the predicted values for all galaxies brighter than $M_{UV}=-15$ at 
that redshift from the theoretical model. Violet points show the values for real galaxies 
in the CANDELS and HUDF fields as inferred by Grazian et al. (A\&A submitted), with 
green points showing the observed medians in each UV bin. The black line shows 
our best-fit theoretical power-law relation, while the blue line 
shows the relation previously inferred from data 
by \citet{gonzalez2011} at $z=4$ (and applied unchanged at higher redshifts).}
\label{m2l} 
\end{figure*}

\section{Stellar mass function and density}

Having checked that our fiducial model predicts the correct evolving 
luminosity function for high-$z$ galaxies, we now use it to study the 
evolving stellar mass function (SMF). The measurement of the SMF at 
extreme redshifts from observations has remained difficult because 
a robust estimate of $M_*$ ideally requires rest-frame near-infrared 
data. In the absence of such data of the required depth, various indirect approaches 
can be used to get an estimate the SMF at a given $z$: 
(a) scaling the HMF with a constant factor assuming a certain $M_*-M_h$ relation, 
(b) scaling the theoretical UV LF assuming a $M_*-M_{UV}$ relation, 
(c) convolving the best-fit functional form (Schechter or DPL) of the 
observed UV LF with a mass-to-light (M/L) ratio including its scatter 
\citep[e.g.][]{gonzalez2011}, and (d) once stellar masses have been derived by 
SED fitting to broad-band colours (albeit without the benefit of 
deep rest-frame near-infrared data), binning up the observed masses can 
yield an estimate of the SMF \citep[e.g.][]{stark2009, labbe2010a}. This last approach is most similar to theoretical models that produce the 
SMF by binning galaxies on the basis of their 
stellar mass \citep[e.g.][]{nagamine2010, dayal2012, 
dayal2013, hutter2014}, albeit with the data the completeness of the 
result SMF is inevitably limited by the depth of the data and the wavelength 
of the selection band. We present the SMFs 
obtained by using all of the above four approaches in Sec. \ref{smf} below.  

\subsection{Mass-to-light relation through time}
\label{sec_m2l}

\begin{figure*} 
   \center{\includegraphics[scale=0.85]{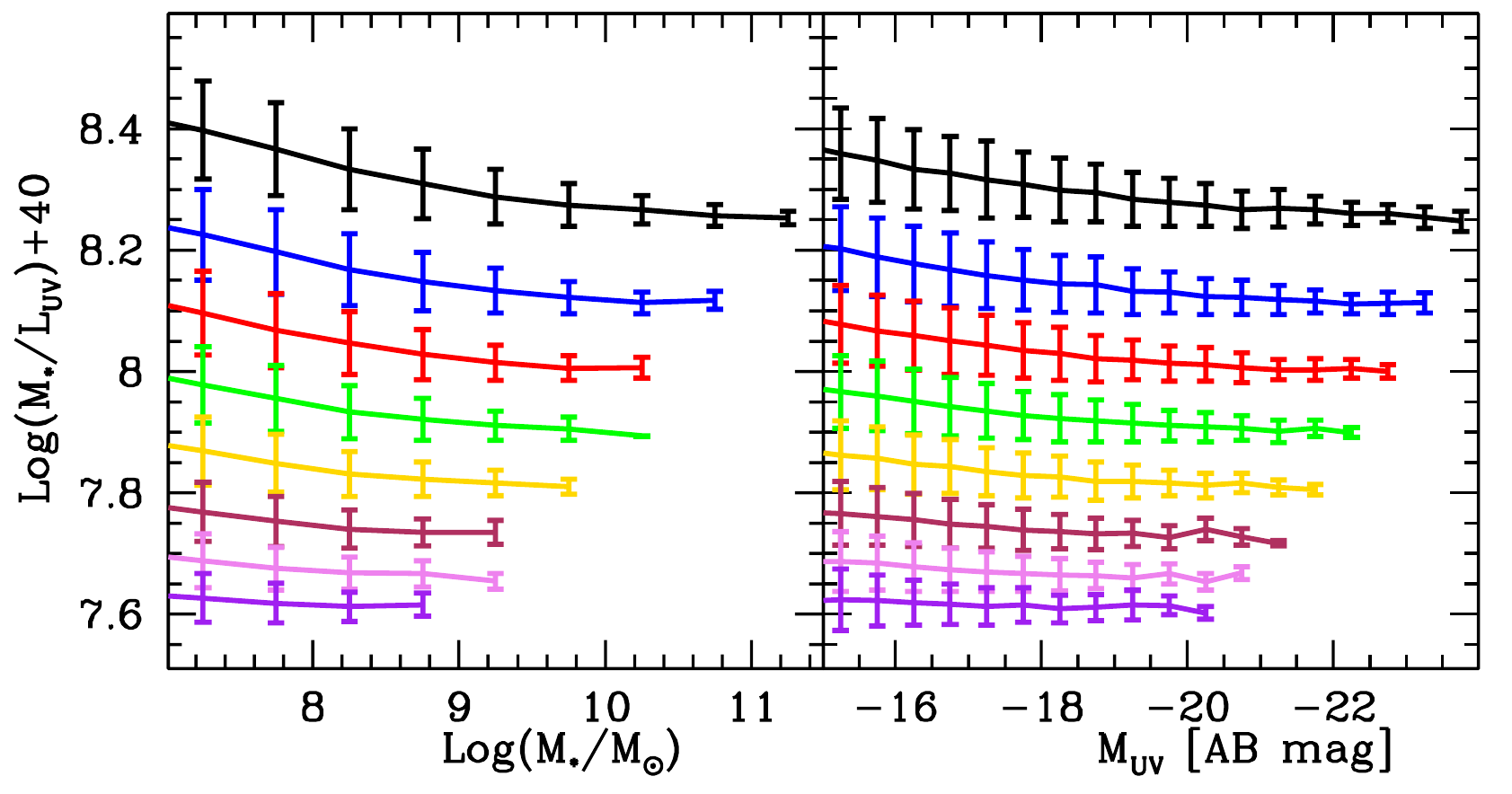}}
  \caption{The theoretical mass-to-light ratio as a function of  stellar mass (left panel) and UV absolute 
magnitude (right panel) for $z \simeq 5-12$ (from top to bottom); $M_*$ and $L_{UV}$ are in units of 
$\Msun$ and  ${\rm erg/s/\AA}$, respectively and we show this ratio arbitrarily scaled up by a factor 
of $10^{40}$.}
\label{m2l2} 
\end{figure*}

We start from the M/L relation that links the total stellar mass $M_*$ and the 
UV luminosity $L_{UV}$, as obtained from our model. As noted in Eqn. \ref{luv}, 
the UV luminosity from a burst of star formation declines rapidly with time 
as $L_{UV} \propto t^{-1.3}$. This implies that the UV luminosity of a starburst 
galaxy is typically dominated by the most recent burst. Combining this with the 
fact that increasingly massive galaxies have larger gas masses available 
for instantaneous star formation and have built up a larger stellar 
mass over their history, we would expect $L_{UV}$ to scale with $M_*$. 
As shown in Fig. \ref{m2l}, this is indeed the behaviour shown by our model. 
We find that the $M_*-M_{UV}$ relation evolves smoothly with redshift with the form
\begin{equation}
\log\,  M_* = \gamma M_{UV} + \delta,
\end{equation}
where the slope of the relation has a constant value of $\gamma \simeq -0.38$ 
at all $z=5$ to $12$. However, the zero-point of the relation changes with redshift such that
\begin{equation}
\delta = -0.13\, z + 2.4,
\label{norm}
\end{equation}
over the redshift range $z = 5 - 12$. Thus, the normalisation of the $M/L$ relation decreases 
by about $0.4$ dex for an increase in redshift by $\Delta z=3$; in other words, galaxies 
of a given luminosity are associated with lower $M_*$ (or $M_h$) values with increasing redshift. As shown in Sec. \ref{sec_phy}, this increase in luminosity arises because galaxies form stars faster with increasing redshift, given the shorter cosmic time available for them to assemble a certain halo mass. 
 
As shown in Fig. \ref{m2l}, the mass-to-light relation recently inferred from 
the CANDELS and HUDF data by Grazian et al. (in preparation; also Duncan et al., in preparation) differs considerably to that 
previously deduced by \citet{gonzalez2011}. Grazian et al. 
find $\gamma = -0.4$ and $\delta = 1.6$ at $3.5 < z <4.5$, which is in 
excellent agreement with our model predictions at $z=5$ as shown in Fig. \ref{m2l}; a combination of observational factors including low galaxy numbers and an increase in the errors associated with the SED fitting to the broad band colours of faint objects ($M_{UV} \gsim -17$) probably leads to the slight mismatch between their observational and our theoretical values at the faint end (A. Grazian, Private communication).
However, the slope ($-0.68$) and normalization ($-4.51$) inferred by 
\citet{gonzalez2011} at $z=4$ (and applied to redshifts as high as $z=7$) are very 
different both from our model and from the new Grazian et al. results. 

 \begin{figure*} 
   \center{\includegraphics[scale=0.9]{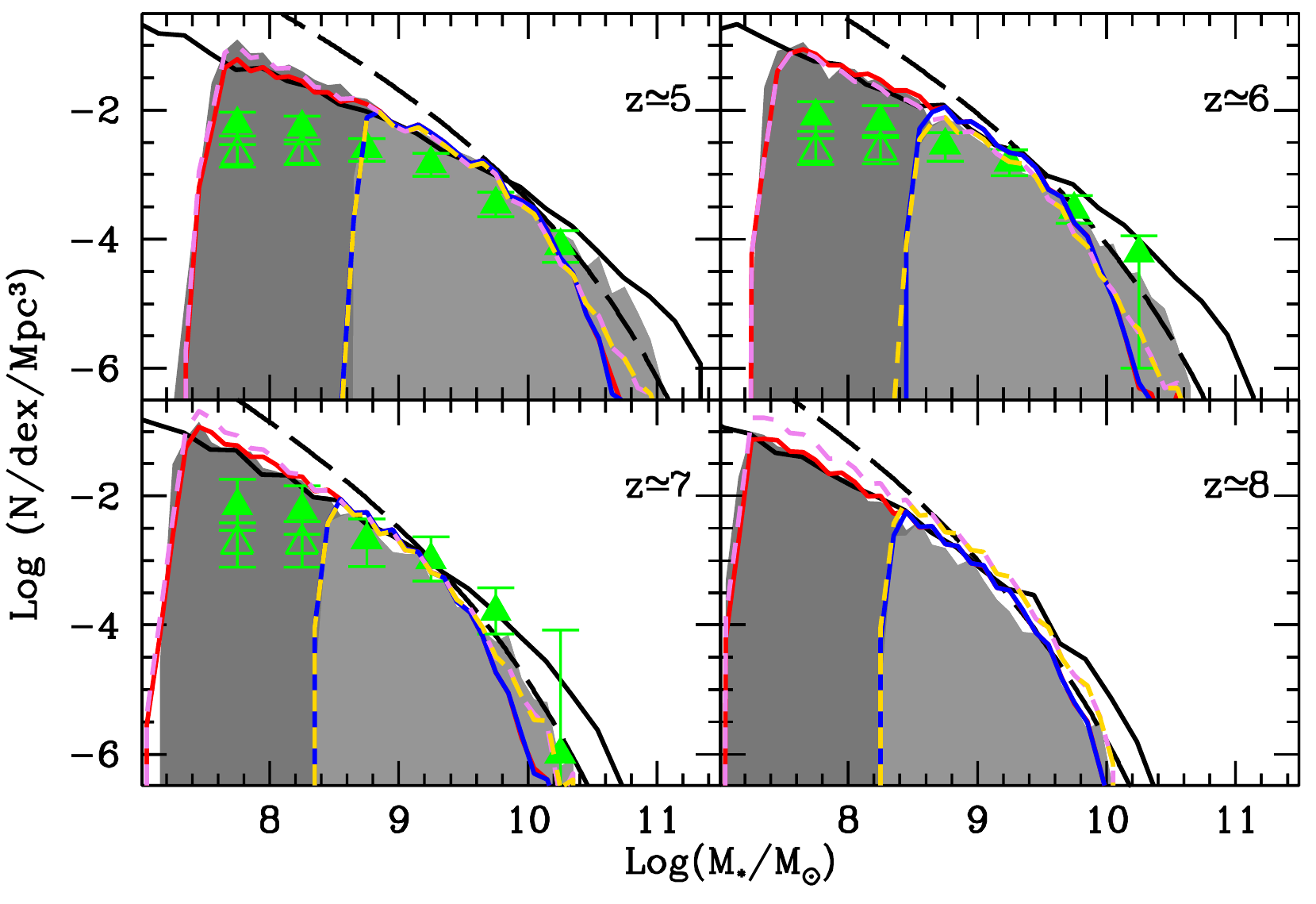}} 
  \caption{The evolving stellar mass function (SMF) for the redshift marked 
in each panel. In each panel, the dark (light) shaded regions show the SMF obtained by 
directly binning model galaxies brighter than $M_{UV}-15$ and $-18$, respectively. 
The solid and dashed black lines have been obtained by scaling the theoretical UV 
LF (Fig. \ref{fig_lbg_uvlf})and DM Halo mass functions at the appropriate redshift 
to match to the massive-end of the theoretical SMFs. The red (blue) lines show the SMF 
obtained by convolving the theoretical $M/L$ ratio (Fig. \ref{m2l}) with the 
observationally-inferred best-fit Schechter function \citep{mclure2009, mclure2013} 
for galaxies brighter than $M_{UV} = -15$ and $-18$, respectively. The violet (gold) lines 
show the SMF obtained by convolving the theoretical $M/L$ ratio with the 
observationally-inferred best-fit double power law (DPL) function \citep{bowler2014} 
for galaxies brighter than $M_{UV} = -15$ and $-18$, respectively; we have used the $z=7$ 
DPL at $z=8$. Filled (empty) points show the corrected (uncorrected) 
stellar mass functions inferred observationally by \citet{gonzalez2011}. }
\label{fig_smf} 
\end{figure*}

The reason for the change in the mass-to-light relation inferred from the observations 
is not completely clear, but is almost certainly in part due to the improved 
near-infrared data ({\it HST} WFC3/IR $Y_{105}$, $J_{125}$, $H_{160}$, and VLT Hawk-I $K$-band)
now available in the relevant deep survey fields. For example, while \citet{gonzalez2011} 
selected galaxies as LBGs based on their UV colours, Grazian et al. have been able to use $H_{160}$ 
as the primary selection band, and even for those galaxies which would still be easily selected 
as LBGs have been able to use near-infrared detections to better constrain ages and hence stellar 
masses (where previously only near-infrared upper limits were available for many objects). 
Whatever the origin of this change, it is clear that new Grazian et al. results are in better 
agreement with the predicted mass-to-light relation which results from our model (with the 
median stellar masses inferred by \citet{gonzalez2011} being apparently 
under-estimated by about an order of magnitude). 
As shown in fig. \ref{m2l}, while the median values inferred by \citet{gonzalez2011} 
slowly converge towards the values found here (and by Grazian et al.) at brighter magnitudes, the 
steeper slope of their $M/L$ relation inevitably affects the inferred SMF, as discuss below 
in Sec. \ref{smf}.

We also show our $M/L$ ratios as a function of $M_*$ and $M_{UV}$ 
in Fig. \ref{m2l2}. As discussed above, galaxies of a given mass assemble 
faster with increasing redshift, i.e. they show higher luminosities for a 
given $M_*$, leading to a decrease in the amplitude of the $M/L$ ratio. As expected 
from the evolution of this relation (see Eqn. \ref{norm}), the $M/L$ ratio decreases 
by about 0.3-0.4 dex for an increase in redshift of $\Delta z =3$, both as a function 
of $M_*$ and $M_{UV}$. Further, while the $M/L$ ratio is flat for $z \geq 9$, 
it becomes a decreasing function of luminosity at later times. This is because the 
rate at which massive galaxies build up and form stars increases with time 
as they gain progressively larger gas masses (both from accretion and mergers), 
leading to a smaller $M/L$ than low-mass systems (see Fig. \ref{assembly}).

\subsection{Building stellar mass functions}
\label{smf}
We now show the evolving SMF obtained directly from our model, and 
compare it to those obtained by scaling the UV LF/HMF, and convolving functional fits 
(Schechter function and double power-law) to the UV LF with our $M/L$ 
ratios (including associated scatter) as shown in Fig. \ref{fig_smf}.
The evolving (fiducial) SMF is obtained by binning the number of theoretical 
galaxies in $M_*$ bins at $z = 5-8$ for galaxies brighter than $M_{UV}=-18$ 
(the approximate limit of current observations) and extending to 
magnitudes as faint as $-15$ (as expected to be detectable by future instruments such as the {\it JWST}. 
Firstly, our model predicts that the faint-end of the SMF continues to rise to masses as 
low as $M_*=10^7 \Msun$ at $z =8$. This lower limit increases to 
$M_* \simeq 10^{7.8} \Msun$ at $z = 5$ as galaxies typically become more massive. 

We now compare the SMF from our model to that built by \citet{gonzalez2011} 
integrating down to $M_{UV}=-18$. Starting with the low-mass end, we find the 
theoretical SMF is shifted towards higher $M_*$ values (by about an order-of-magnitude) 
at all $z=5-8$, since \citet{gonzalez2011} have under-estimated the 
median $M/L$ ratio at this end (see sec. \ref{sec_m2l}). 
At the massive-end, our SMF samples much higher $M_*$ 
values since we integrate up to magnitudes as high as $M_{UV} = 
(-23.75,-22.5)$ at $z = (5,8)$, compared to the value of $M_{UV}=-21$ used 
by \citet{gonzalez2011}. However, the shape and amplitude of the theoretical 
and observationally-inferred SMFs are quite similar in the mass range 
$10^{8.4-10} \Msun$ due to the underlying function being compatible with a Schechter 
function in both cases; we remind the reader that while such functional 
shape arises naturally in our model, \citet{gonzalez2011} have obtained 
the SMF by explicitly convolving a Schechter function with their $M/L$ ratio. 

We now present theoretical SMFs obtained by convolving the $M/L$ ratio with 
two different functional forms of the UV LF: the Schechter function 
\citep[parameters from][]{mclure2009, mclure2013} and the DPL \citep[parameters from][]{bowler2014}; 
we use the same DPL parameters at $z=8$ as at $z=7$. For each function, 
we sample the $M_*$ value in each $M_{UV}$ bin shown in Fig. \ref{m2l} 
assuming a Gaussian error distribution to build the boot-strapped SMFs 
shown in Fig. \ref{fig_smf}. Starting at the massive-end, we find that the fiducial 
SMF is in better agreement with that obtained from the DPL at $z=7$, while the 
SMF from the Schechter function under-estimates the SMF. This is as expected 
given that the fiducial UV LF is compatible with the slow drop-off shown by a 
double power-law for $z=7$ at the bright-end (Sec. \ref{sec_phy}). Mirroring the 
fiducial UV LF that is over-estimated with respect to observations at $z=5,6$, the fiducial 
SMF is also over-estimated at the massive-end compared to both the Schechter 
function and the DPL. The UV LF is compatible with both the Schechter function 
and the DPL at the faint end, leading to the less-massive end of the fiducial SMF 
being in reasonable agreement with the SMFs from both these functions; as expected, the SMF inferred by \citet{gonzalez2011} is flatter than that 
obtained by either function given the different $M/L$ provided by this work. 

Interestingly, while the SMF obtained by arbitrarily scaling the UV LF matches 
the fiducial SMF at the faint end, it over-predicts the mass (by about 0.15 dex) 
for a given number density for $M_* \geq 10^{9.5} \Msun$. As shown in Fig. \ref{m2l2}, 
this is because the $M/L$ ratios of galaxies decrease with increasing $M_*$ (or $M_{UV}$), 
i.e. the SMF would be {\it over-estimated} at the bright-end assuming a constant $M/L$ ratio across the whole UV LF.

As for the SMF obtained by scaling the HMF to match the bright-end, we find it over-estimates 
the faint-end. This is as expected, given that the galaxies at the faint-end are feedback-limited 
due to their progenitors having ejected all their gas; 
this reduces their $M_*$ (and $M_{UV}$) value below that for larger galaxies. 

Interestingly, while the HMF scaling required to reproduce the UV LF (see Table 1 and Fig. \ref{fig_lbg_uvlf}) is a function of redshift, the HMF scaling required to reproduce the bright end of the SMF (Fig. \ref{fig_smf}) is independent of redshift and has the physically reasonable value of $M_* = 0.016 M_h$. Essentially these two figures demonstrate that our fiducial model delivers the correct LF shape which cannot be produced by simple scaling of the HMF, while at the same time producing a stellar mass function which, at the high mass end, does indeed mirror the shape of the HMF.

To summarize, we find that assuming a constant $M/L$ ratio across the whole UV 
LF would lead to an over-estimation of the massive end of the SMF, while a scaled 
HMF overpredicts the faint end. A good match to the fiducial SMF can then be obtained 
from an {\it $M_*$ that traces the UV luminosity at the low-mass end and the halo mass at the high-mass end}. 

\begin{figure}
\center{\includegraphics[scale=0.49]{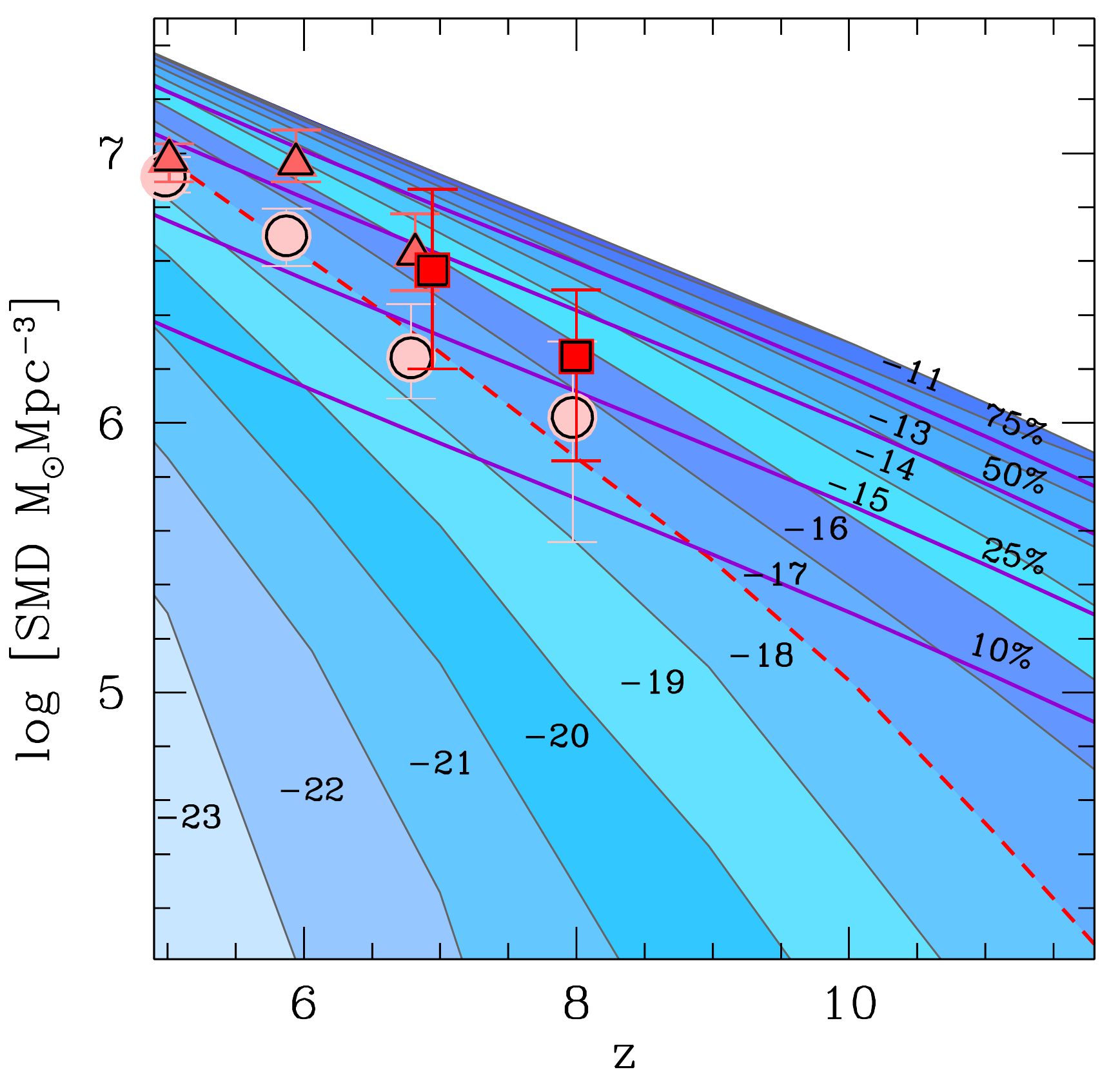}}
\caption{The stellar mass density ($SMD$) as a function of redshift. The 
different coloured contours show the contribution to the SMD from galaxies 
brighter than the magnitude value marked in the contour. The solid purple lines 
show fractions of the SMD to allow estimates of the magnitude limits which reveal the 
galaxies providing 10\%, 25\%, 50\% and 75\% of the total SMD at any redshift. 
The dashed red line shows the SMD from galaxies that have already been detected ($M_{UV} \lsim -18)$ to 
allow comparison with the data points: \citet[filled triangles]{gonzalez2011}, 
\citet[filled circles]{stark2013} and \citet[filled squares]{labbe2010a,labbe2010b}. 
Our model predicts that most of the stellar mass in the Universe at $5 \leq z \leq 12$ 
is locked up in systems too faint to be detected by the {\it HST}, but deep surveys with the
{\it JWST} should reveal over a half 
(a fourth) of the total stellar mass in the Universe at redshifts as high at $z \simeq 5$ $(9.5)$.}
\label{smd} 
\end{figure}

\subsection{Cosmic stellar mass census}
Our model also yields the stellar mass density (SMD; stellar mass per unit comoving volume) 
as a function of redshift, which can be directly compared to observations. 
Since both the halo mass function and the host halos of equally-luminous galaxies 
shift to progressively lower masses with increasing redshift, a comparison of the total 
theoretical SMD with observations is clearly a non-trivial test of our model. Encouragingly, the 
results of the model are in extremely good agreement with observationally inferred SMD values as shown 
in Fig.\,9. Conducting a census of the total stellar mass we find that, as a result of 
their enormous numbers, small, faint galaxies contain most of the stellar mass 
in the Universe at $5 \leq z \leq 12$. Indeed, galaxies brighter than current observational 
limits ($ M_{UV} \leq -18$) contain about $50\%$ of the total stellar mass at $z \simeq 5$. 
This value then steadily decreases with redshift such that observed galaxies contain 
a quarter of the total stellar mass at $z \simeq 6.5$ and only $10\%$ of the 
total stellar mass at $z \simeq 9$ (i.e. at redshifts $z \geq 6.5$, 
most of the stellar mass of the Universe is locked up in galaxies too small to have been detected so far). 
The next generation of space instruments such as the {\it JWST}, along with future developments 
in the use of gravitational lensing (e.g. in the Frontier Fields), will play an important role 
in revealing about half (a quarter) of the stellar mass in the Universe up to $z \simeq 8$ ($z \simeq 9.5$). 

\section{Conclusions}
Recent deep {\it HST} and ground-based surveys have enabled the construction of 
statistically-significant samples of galaxies at $z \geq 5$, providing key new information 
on the evolving galaxy ultraviolet luminosity function (UV LF), mass-to-light ratios ($M/L$), stellar mass functions (SMF) and the integrated 
stellar mass Density (SMD) out to $z \simeq 10$. In this study we have endeavoured to isolate the essential physics driving early galaxy 
formation/evolution by building a very simple semi-analytical model wherein DM merger trees are implemented 
with the key physics of star formation, SN feedback and the resulting gas ejection, 
and the growth of progressively more massive systems (via halo mergers and gas accretion). 
In the spirit of maintaining simplicity, our model utilises a total of two 
{\it redshift- and mass-independent free-parameters}: the star formation efficiency threshold, $f_*$, and 
the fraction of SN energy that drives winds, $f_w$. Our model is based on the single premise 
that any galaxy can form stars with a maximal {\it effective efficiency ($f_*^{eff}$) that provides 
enough energy to expel all the remaining gas, quenching further star formation}. The value of 
$f_*^{eff} = min[f_*,f_*^{ej}]$ where $f_*^{ej}$ is the star-formation efficiency required to 
eject all gas from a galaxy. In this model, low-mass galaxies form stars with an effective 
efficiency that is sufficient to eject all gas from a galaxy $f_*^{eff} = f_*^{ej}$, 
while massive systems can form stars with a larger fixed efficiency value of $f_*^{eff} = f_*$. 

This simple model reproduces both the slope and amplitude of the evolving UV LF at $z=5$ to $z=10$ 
with parameter values $f_* = 0.03$ and $f_w = 0.1$ (i.e. a model wherein at most 3\% of the 
gas can form into stars and 10\% of SNe energy is converted into kinetic form to drive outflows). 
Although the Halo Mass Function (HMF) scaled to match the knee of the observed UV LF 
shifts to progressively lower masses at high $z$, our model correctly reproduces the evolving 
faint-end slope ($\alpha$) of the UV LF; the physical explanation is the faster assembly of galaxies with increasing redshift. Additionally, the model naturally predicts that 
the bright-end slope of the UV LF is somewhat flatter than the steep exponential drop-off 
provided by the Schechter function, and is better described as either following a double 
power-law \citet{bowler2014}, or by the high-mass shape of the HMF. 

Galaxies with $M_{UV} \leq -15$ are hosted in halos with mass $M_h \geq 10^{9.5} (10^9) \Msun$ 
for $z=5-8\, (12)$, a mass scale above the feedback-limited regime at these redshifts. 
Hence, the faint-end of the UV LF lies below the HMF because the progenitors of these 
galaxies have ejected all their gas, resulting in dry mergers. These systems are therefore 
{\it fuel-starved}, rather than feedback-limited. We also show that the physics driving 
galaxy growth is halo mass-dependent: while the smallest halos build up their gas 
(and stellar) mass by accreting gas from the IGM, the bulk of the gas powering star formation 
in the largest halos is brought in by merging progenitors. From our model we have provided 
a simple functional form for the redshift evolution of the faint-end slope, 
$\alpha = -1.75 \log \,z -0.52$, a key ingredient for reionization 
calculations since faint galaxies provide the bulk of hydrogen ionizing photons ($E \ge 1$ Ryd).

The stellar mass-to-light relation from our model is well-fit by the functional form $\log\,  M_* = -0.38 M_{UV} -0.13\, z + 2.4$. This relation is in excellent agreement with the recent observational results 
obtained by Grazian et al. (A\&A submitted), but is significantly flatter (i.e. higher average masses at low luminosities) 
than the relation previously deduced from observations of LBGs by \citet{gonzalez2010}. We also 
show the SMFs obtained by (a) binning up $M_*$ from the fiducial model, (b) convolving the $M/L$ 
relation with a Schechter function and DPL, (c) scaling the fiducial UV LF, and (d) 
scaling the HMF. A good match to the fiducial SMF can be obtained by using a 
$M_*$ tracing the UV luminosity at the faint-end and the HMF at the high-mass end. 

The census of the cosmic SMD implies that while 50\% of the SMD is in detected 
LBGs at $z \simeq 5$, this fraction drops steadily to 10\% at $z \simeq 9$. 
While the next generation of instruments such as the {\it JWST} should be capable 
of revealing up to 25\% of the SMD at $z \simeq 9.5$, most of the SMD at high-$z$ 
is predicted to be locked up in galaxies that will likely remain ``invisible" for the immediate 
future.

We end with a few caveats. First, it appears that the fiducial model slightly over-predicts the 
bright-end of the UV LF at $z =5,6$. This could either arise due to physical 
effects that have been ignored in order to maintain simplicity (e.g. dust attenuation, 
AGN feedback, mass quenching) which may become increasingly important with reducing redshift, 
or due to current data limitations at the brightest UV luminosities (a situation that should be resolved soon
by Bowler et al. in preparation). 
Second, our model assumes a maximally-efficient feedback scenario wherein SN-powered 
kinetic winds in small galaxies can sweep up all the gas and eject it out of the system.  
Although this is justifiable based on energy balance arguments, it might be possible that 
a fraction of the gas remains within the halo potential well, albeit in a heated/dynamically 
perturbed state not allowing stars to readily form \citep{Mori2002, Fangano2007}. Third, we have ignored feedback mechanisms including (i) the energy injected by massive stars before the onset of SN explosions \citep{hopkins2011, stinson2013} and, (ii) the photo-evaporation of gas on the outskirts of galaxies due to a UV background (UVB) created by reionization. Point (i) is justified by our assumption of instantaneous star formation and ISM SN energy injection. However, making the simplifying assumption that stellar feedback only adds to the total energy injected into the ISM gas (Eqn. 1), $f_*^{ej}$ can be left unchanged by decreasing the value of $f_w$ by the same amount. As for (ii), we show that SN feedback always dominates over the effect of the UVB \citep{dayal2014b}; hence including the decrease in the baryon fraction due to a UVB would leave our results unchanged. However, these details of feedback are highly model-dependent and progress can be made only 
by comparing such predictions with actual data as those imminently expected from the 
Frontier Fields and ground-based observatories (Yue et al. 2014, in prep.)

In the future, our aim is to include the effects of AGB dust, AGN feedback and the UVB created by both Hydrogen and helium reioniozation. These are some of the main physical effects whose inclusion will allow our model to be extended all the way to $z=0$.
\section*{Acknowledgments} 
JSD and PD acknowledge the support of the European Research Council via the award of an Advanced Grant. 
JSD also acknowledges the support of the Royal Society via a Wolfson Research Merit award, and the contribution of 
the EC FP7 SPACE project ASTRODEEP (Ref. No: 312725). 
The authors thank A. Grazian and co-authors for allowing 
us to use their results and for their constructive comments.
PD thanks A. Mazumdar and the anonymous referee for their insightful comments.


\bibliographystyle{mn2e}
\bibliography{fb}

\begin{thebibliography}{88}
\expandafter\ifx\csname natexlab\endcsname\relax\def\natexlab#1{#1}\fi

\bibitem[{{Aguirre} {et~al}\mbox{.}(2001){Aguirre}, {Hernquist}, {Schaye},
  {Katz}, {Weinberg}, \& {Gardner}}]{aguirre2001}
{Aguirre} A., {Hernquist} L., {Schaye} J., {Katz} N., {Weinberg} D.~H.,
  {Gardner} J., 2001, \apj, 561, 521

\bibitem[{{Baugh}(2006)}]{baugh2006}
{Baugh} C.~M., 2006, Reports on Progress in Physics, 69, 3101

\bibitem[{{Benson}(2012)}]{benson2012}
{Benson} A.~J., 2012, \na, 17, 175

\bibitem[{{Benson} {et~al}\mbox{.}(2003){Benson}, {Bower}, {Frenk}, {Lacey},
  {Baugh}, \& {Cole}}]{benson2003}
{Benson} A.~J., {Bower} R.~G., {Frenk} C.~S., {Lacey} C.~G., {Baugh} C.~M.,
  {Cole} S., 2003, \apj, 599, 38

\bibitem[{{Bouwens} {et~al}\mbox{.}(2007){Bouwens}, {Illingworth}, {Franx}, \&
  {Ford}}]{bouwens2007}
{Bouwens} R.~J., {Illingworth} G.~D., {Franx} M., {Ford} H., 2007, \apj, 670,
  928

\bibitem[{{Bouwens} {et~al}\mbox{.}(2010{\natexlab{a}}){Bouwens},
  {Illingworth}, {Gonz{\'a}lez}, {Labb{\'e}}, {Franx}, {Conselice},
  {Blakeslee}, {van Dokkum}, {Holden}, {Magee}, {Marchesini}, \&
  {Zheng}}]{bouwens2010a}
{Bouwens} R.~J. {et~al.}, 2010{\natexlab{a}}, \apj, 725, 1587

\bibitem[{{Bouwens} {et~al}\mbox{.}(2011){Bouwens}, {Illingworth}, {Oesch},
  {Labb{\'e}}, {Trenti}, {van Dokkum}, {Franx}, {Stiavelli}, {Carollo},
  {Magee}, \& {Gonzalez}}]{bouwens2011b}
{Bouwens} R.~J. {et~al.}, 2011, \apj, 737, 90

\bibitem[{{Bouwens} {et~al}\mbox{.}(2013){Bouwens}, {Illingworth}, {Oesch},
  {Labbe}, {van Dokkum}, {Trenti}, {Franx}, {Smit}, {Gonzalez}, \&
  {Magee}}]{bouwens2013}
{Bouwens} R.~J. {et~al.}, 2013, ArXiv:1306.2950

\bibitem[{{Bouwens} {et~al}\mbox{.}(2014){Bouwens}, {Illingworth}, {Oesch},
  {Trenti}, {Labbe'}, {Bradley}, {Carollo}, {van Dokkum}, {Gonzalez},
  {Holwerda}, {Franx}, {Spitler}, {Smit}, \& {Magee}}]{bouwens2014}
{Bouwens} R.~J. {et~al.}, 2014, ArXiv:1403.4295

\bibitem[{{Bouwens} {et~al}\mbox{.}(2010{\natexlab{b}}){Bouwens},
  {Illingworth}, {Oesch}, {Trenti}, {Stiavelli}, {Carollo}, {Franx}, {van
  Dokkum}, {Labb{\'e}}, \& {Magee}}]{bouwens2010b}
{Bouwens} R.~J. {et~al.}, 2010{\natexlab{b}}, \apjl, 708, L69

\bibitem[{{Bower}(1991)}]{bower1991}
{Bower} R.~G., 1991, \mnras, 248, 332

\bibitem[{{Bowler} {et~al}\mbox{.}(2012){Bowler}, {Dunlop}, {McLure},
  {McCracken}, {Milvang-Jensen}, {Furusawa}, {Fynbo}, {Le Fevre}, {Holt},
  {Ideue}, {Ihara}, {Rogers}, \& {Taniguchi}}]{bowler2012}
{Bowler} R.~A.~A. {et~al.}, 2012, ArXiv e-prints

\bibitem[{{Bowler} {et~al}\mbox{.}(2014){Bowler}, {Dunlop}, {McLure}, {Rogers},
  {McCracken}, {Milvang-Jensen}, {Furusawa}, {Fynbo}, {Taniguchi}, {Afonso},
  {Bremer}, \& {Le F{\`e}vre}}]{bowler2014}
{Bowler} R.~A.~A. {et~al.}, 2014, \mnras, 440, 2810

\bibitem[{{Bradley} {et~al}\mbox{.}(2012){Bradley}, {Trenti}, {Oesch},
  {Stiavelli}, {Treu}, {Bouwens}, {Shull}, {Holwerda}, \&
  {Pirzkal}}]{bradley2012}
{Bradley} L.~D. {et~al.}, 2012, \apj, 760, 108

\bibitem[{{Bradley} {et~al}\mbox{.}(2013){Bradley}, {Zitrin}, {Coe}, {Bouwens},
  {Postman}, {Balestra}, {Grillo}, {Monna}, {Rosati}, {Seitz}, {Host}, {Lemze},
  {Moustakas}, {Moustakas}, {Shu}, {Zheng}, {Broadhurst}, {Carrasco}, {Jouvel},
  {Koekemoer}, {.~Medezinski}, {Meneghetti}, {Nonino}, {Smit}, {Umetsu},
  {Bartelmann}, {Benitez}, {Donahue}, {Ford}, {Infante}, {Jimenez-Teja},
  {Kelson}, {Lahav}, {Maoz}, {Melchior}, {Merten}, \& {Molino}}]{bradley2013}
{Bradley} L.~D. {et~al.}, 2013, ArXiv:1308.1692

\bibitem[{{Castellano} {et~al}\mbox{.}(2010){Castellano}, {Fontana}, {Paris},
  {Grazian}, {Pentericci}, {Boutsia}, {Santini}, {Testa}, {Dickinson},
  {Giavalisco}, {Bouwens}, {Cuby}, {Mannucci}, {Cl{\'e}ment}, {Cristiani},
  {Fiore}, {Gallozzi}, {Giallongo}, {Maiolino}, {Menci}, {Moorwood}, {Nonino},
  {Renzini}, {Rosati}, {Salimbeni}, \& {Vanzella}}]{castellano2010}
{Castellano} M. {et~al.}, 2010, \aap, 524, A28

\bibitem[{{Choudhury} \& {Ferrara}(2007)}]{choudhury-ferrara2007}
{Choudhury} T.~R., {Ferrara} A., 2007, \mnras, 380, L6

\bibitem[{{Coe} {et~al}\mbox{.}(2013){Coe}, {Zitrin}, {Carrasco}, {Shu},
  {Zheng}, {Postman}, {Bradley}, {Koekemoer}, {Bouwens}, {Broadhurst}, {Monna},
  {Host}, {Moustakas}, {Ford}, {Moustakas}, {van der Wel}, {Donahue}, {Rodney},
  {Ben{\'{\i}}tez}, {Jouvel}, {Seitz}, {Kelson}, \& {Rosati}}]{coe2013}
{Coe} D. {et~al.}, 2013, \apj, 762, 32

\bibitem[{{Cole}(1991)}]{cole1991}
{Cole} S., 1991, \apj, 367, 45

\bibitem[{{Cole} {et~al}\mbox{.}(1994){Cole}, {Aragon-Salamanca}, {Frenk},
  {Navarro}, \& {Zepf}}]{cole1994}
{Cole} S., {Aragon-Salamanca} A., {Frenk} C.~S., {Navarro} J.~F., {Zepf} S.~E.,
  1994, \mnras, 271, 781

\bibitem[{{Croton} {et~al}\mbox{.}(2006){Croton}, {Springel}, {White}, {De
  Lucia}, {Frenk}, {Gao}, {Jenkins}, {Kauffmann}, {Navarro}, \&
  {Yoshida}}]{croton2006}
{Croton} D.~J. {et~al.}, 2006, \mnras, 365, 11

\bibitem[{{Dayal} {et~al}\mbox{.}(2013){Dayal}, {Dunlop}, {Maio}, \&
  {Ciardi}}]{dayal2013}
{Dayal} P., {Dunlop} J.~S., {Maio} U., {Ciardi} B., 2013, \mnras, 434, 1486

\bibitem[{{Dayal} \& {Ferrara}(2012)}]{dayal2012}
{Dayal} P., {Ferrara} A., 2012, \mnras, 421, 2568

\bibitem[{{Dayal} {et~al}\mbox{.}(2010){Dayal}, {Ferrara}, \&
  {Saro}}]{dayal2010a}
{Dayal} P., {Ferrara} A., {Saro} A., 2010, \mnras, 402, 1449

\bibitem[{{Dayal} {et~al}\mbox{.}(2009){Dayal}, {Ferrara}, {Saro},
  {Salvaterra}, {Borgani}, \& {Tornatore}}]{dayal2009}
{Dayal} P., {Ferrara} A., {Saro} A., {Salvaterra} R., {Borgani} S., {Tornatore}
  L., 2009, \mnras, 400, 2000

\bibitem[{{Dayal} {et~al}\mbox{.}(2014){Dayal}, {Mesinger}, \&
  {Pacucci}}]{dayal2014b}
{Dayal} P., {Mesinger} A., {Pacucci} F., 2014, ArXiv e-prints

\bibitem[{{De Lucia} {et~al}\mbox{.}(2010){De Lucia}, {Boylan-Kolchin},
  {Benson}, {Fontanot}, \& {Monaco}}]{delucia2010}
{De Lucia} G., {Boylan-Kolchin} M., {Benson} A.~J., {Fontanot} F., {Monaco} P.,
  2010, \mnras, 406, 1533

\bibitem[{{Dubinski} \& {Carlberg}(1991)}]{dubinski-carlberg1991}
{Dubinski} J., {Carlberg} R.~G., 1991, \apj, 378, 496

\bibitem[{{Dunlop} {et~al}\mbox{.}(2012){Dunlop}, {McLure}, {Robertson},
  {Ellis}, {Stark}, {Cirasuolo}, \& {de Ravel}}]{dunlop2012}
{Dunlop} J.~S., {McLure} R.~J., {Robertson} B.~E., {Ellis} R.~S., {Stark}
  D.~P., {Cirasuolo} M., {de Ravel} L., 2012, \mnras, 420, 901

\bibitem[{{Dunlop} {et~al}\mbox{.}(2013){Dunlop}, {Rogers}, {McLure}, {Ellis},
  {Robertson}, {Koekemoer}, {Dayal}, {Curtis-Lake}, {Wild}, {Charlot},
  {Bowler}, {Schenker}, {Ouchi}, {Ono}, {Cirasuolo}, {Furlanetto}, {Stark},
  {Targett}, \& {Schneider}}]{dunlop2013}
{Dunlop} J.~S. {et~al.}, 2013, \mnras, 432, 3520

\bibitem[{{Ellis} {et~al}\mbox{.}(2013){Ellis}, {McLure}, {Dunlop},
  {Robertson}, {Ono}, {Schenker}, {Koekemoer}, {Bowler}, {Ouchi}, {Rogers},
  {Curtis-Lake}, {Schneider}, {Charlot}, {Stark}, {Furlanetto}, \&
  {Cirasuolo}}]{ellis2013}
{Ellis} R.~S. {et~al.}, 2013, \apjl, 763, L7

\bibitem[{{Fangano} {et~al}\mbox{.}(2007){Fangano}, {Ferrara}, \&
  {Richter}}]{Fangano2007}
{Fangano} A.~P.~M., {Ferrara} A., {Richter} P., 2007, \mnras, 381, 469

\bibitem[{{Finkelstein} {et~al}\mbox{.}(2012){Finkelstein}, {Papovich},
  {Salmon}, {Finlator}, {Dickinson}, {Ferguson}, {Giavalisco}, {Koekemoer},
  {Reddy}, {Bassett}, {Conselice}, {Dunlop}, {Faber}, {Grogin}, {Hathi},
  {Kocevski}, {Lai}, {Lee}, {McLure}, {Mobasher}, \&
  {Newman}}]{finkelstein2012}
{Finkelstein} S.~L. {et~al.}, 2012, \apj, 756, 164

\bibitem[{{Finlator} {et~al}\mbox{.}(2007){Finlator}, {Dav{\'e}}, \&
  {Oppenheimer}}]{finlator2007}
{Finlator} K., {Dav{\'e}} R., {Oppenheimer} B.~D., 2007, \mnras, 376, 1861

\bibitem[{{Fixsen} {et~al}\mbox{.}(1996){Fixsen}, {Cheng}, {Gales}, {Mather},
  {Shafer}, \& {Wright}}]{fixsen1996}
{Fixsen} D.~J., {Cheng} E.~S., {Gales} J.~M., {Mather} J.~C., {Shafer} R.~A.,
  {Wright} E.~L., 1996, \apj, 473, 576

\bibitem[{{Forero-Romero} {et~al}\mbox{.}(2011){Forero-Romero}, {Yepes},
  {Gottl{\"o}ber}, {Knollmann}, {Cuesta}, \& {Prada}}]{forero2011}
{Forero-Romero} J.~E., {Yepes} G., {Gottl{\"o}ber} S., {Knollmann} S.~R.,
  {Cuesta} A.~J., {Prada} F., 2011, \mnras, 415, 3666

\bibitem[{{Gnedin}(1998)}]{gnedin1998}
{Gnedin} N.~Y., 1998, \mnras, 294, 407

\bibitem[{{Gonz{\'a}lez} {et~al}\mbox{.}(2011){Gonz{\'a}lez}, {Labb{\'e}},
  {Bouwens}, {Illingworth}, {Franx}, \& {Kriek}}]{gonzalez2011}
{Gonz{\'a}lez} V., {Labb{\'e}} I., {Bouwens} R.~J., {Illingworth} G., {Franx}
  M., {Kriek} M., 2011, \apjl, 735, L34

\bibitem[{{Gonz{\'a}lez} {et~al}\mbox{.}(2010){Gonz{\'a}lez}, {Labb{\'e}},
  {Bouwens}, {Illingworth}, {Franx}, {Kriek}, \& {Brammer}}]{gonzalez2010}
{Gonz{\'a}lez} V., {Labb{\'e}} I., {Bouwens} R.~J., {Illingworth} G., {Franx}
  M., {Kriek} M., {Brammer} G.~B., 2010, \apj, 713, 115

\bibitem[{{Greif} {et~al}\mbox{.}(2007){Greif}, {Johnson}, {Bromm}, \&
  {Klessen}}]{greif2007}
{Greif} T.~H., {Johnson} J.~L., {Bromm} V., {Klessen} R.~S., 2007, \apj, 670, 1

\bibitem[{{Hinshaw} {et~al}\mbox{.}(2013){Hinshaw}, {Larson}, {Komatsu},
  {Spergel}, {Bennett}, {Dunkley}, {Nolta}, {Halpern}, {Hill}, {Odegard},
  {Page}, {Smith}, {Weiland}, {Gold}, {Jarosik}, {Kogut}, {Limon}, {Meyer},
  {Tucker}, {Wollack}, \& {Wright}}]{hinshaw2013}
{Hinshaw} G. {et~al.}, 2013, \apjs, 208, 19

\bibitem[{{Hopkins} {et~al}\mbox{.}(2011){Hopkins}, {Quataert}, \&
  {Murray}}]{hopkins2011}
{Hopkins} P.~F., {Quataert} E., {Murray} N., 2011, \mnras, 417, 950

\bibitem[{{Hutter} {et~al}\mbox{.}(2014){Hutter}, {Dayal}, {Partl}, \&
  {M{\"u}ller}}]{hutter2014}
{Hutter} A., {Dayal} P., {Partl} A.~M., {M{\"u}ller} V., 2014, ArXiv:1401.2830

\bibitem[{{Jaacks} {et~al}\mbox{.}(2012){Jaacks}, {Choi}, {Nagamine},
  {Thompson}, \& {Varghese}}]{jaacks2012}
{Jaacks} J., {Choi} J.-H., {Nagamine} K., {Thompson} R., {Varghese} S., 2012,
  \mnras, 420, 1606

\bibitem[{{Klypin} {et~al}\mbox{.}(1999){Klypin}, {Kravtsov}, {Valenzuela}, \&
  {Prada}}]{klypin1999}
{Klypin} A., {Kravtsov} A.~V., {Valenzuela} O., {Prada} F., 1999, \apj, 522, 82

\bibitem[{{Labb{\'e}} {et~al}\mbox{.}(2010{\natexlab{a}}){Labb{\'e}},
  {Gonz{\'a}lez}, {Bouwens}, {Illingworth}, {Franx}, {Trenti}, {Oesch}, {van
  Dokkum}, {Stiavelli}, {Carollo}, {Kriek}, \& {Magee}}]{labbe2010a}
{Labb{\'e}} I. {et~al.}, 2010{\natexlab{a}}, \apjl, 716, L103

\bibitem[{{Labb{\'e}} {et~al}\mbox{.}(2010{\natexlab{b}}){Labb{\'e}},
  {Gonz{\'a}lez}, {Bouwens}, {Illingworth}, {Oesch}, {van Dokkum}, {Carollo},
  {Franx}, {Stiavelli}, {Trenti}, {Magee}, \& {Kriek}}]{labbe2010b}
{Labb{\'e}} I. {et~al.}, 2010{\natexlab{b}}, \apjl, 708, L26

\bibitem[{{Labb{\'e}} {et~al}\mbox{.}(2013){Labb{\'e}}, {Oesch}, {Bouwens},
  {Illingworth}, {Magee}, {Gonz{\'a}lez}, {Carollo}, {Franx}, {Trenti}, {van
  Dokkum}, \& {Stiavelli}}]{labbe2013}
{Labb{\'e}} I. {et~al.}, 2013, \apjl, 777, L19

\bibitem[{{Lacey} \& {Cole}(1993)}]{lacey-cole1993}
{Lacey} C., {Cole} S., 1993, \mnras, 262, 627

\bibitem[{{Lacey} \& {Silk}(1991)}]{lacey-silk1991}
{Lacey} C., {Silk} J., 1991, \apj, 381, 14

\bibitem[{{Lange} {et~al}\mbox{.}(2001){Lange}, {Ade}, {Bock}, {Bond},
  {Borrill}, {Boscaleri}, {Coble}, {Crill}, {de Bernardis}, {Farese},
  {Ferreira}, {Ganga}, {Giacometti}, {Hivon}, {Hristov}, {Iacoangeli}, {Jaffe},
  {Martinis}, {Masi}, {Mauskopf}, {Melchiorri}, {Montroy}, {Netterfield},
  {Pascale}, {Piacentini}, {Pogosyan}, {Prunet}, {Rao}, {Romeo}, {Ruhl},
  {Scaramuzzi}, \& {Sforna}}]{lange2001}
{Lange} A.~E. {et~al.}, 2001, \prd, 63, 042001

\bibitem[{{Leitherer} {et~al}\mbox{.}(1999){Leitherer}, {Schaerer}, {Goldader},
  {Gonz{\'a}lez Delgado}, {Robert}, {Kune}, {de Mello}, {Devost}, \&
  {Heckman}}]{leitherer1999}
{Leitherer} C. {et~al.}, 1999, \apjs, 123, 3

\bibitem[{{Lu} {et~al}\mbox{.}(2013){Lu}, {Wechsler}, {Somerville}, {Croton},
  {Porter}, {Primack}, {Behroozi}, {Ferguson}, {Koo}, {Guo}, {Safarzadeh},
  {Finlator}, {Castellano}, {White}, {Sommariva}, \& {Moody}}]{lu2013}
{Lu} Y. {et~al.}, 2013, ArXiv:1312.3233

\bibitem[{{Mac Low} \& {Ferrara}(1999)}]{maclow-ferrara1999}
{Mac Low} M.-M., {Ferrara} A., 1999, \apj, 513, 142

\bibitem[{{McLure} {et~al}\mbox{.}(2009){McLure}, {Cirasuolo}, {Dunlop},
  {Foucaud}, \& {Almaini}}]{mclure2009}
{McLure} R.~J., {Cirasuolo} M., {Dunlop} J.~S., {Foucaud} S., {Almaini} O.,
  2009, \mnras, 395, 2196

\bibitem[{{McLure} {et~al}\mbox{.}(2013){McLure}, {Dunlop}, {Bowler},
  {Curtis-Lake}, {Schenker}, {Ellis}, {Robertson}, {Koekemoer}, {Rogers},
  {Ono}, {Ouchi}, {Charlot}, {Wild}, {Stark}, {Furlanetto}, {Cirasuolo}, \&
  {Targett}}]{mclure2013}
{McLure} R.~J. {et~al.}, 2013, \mnras, 432, 2696

\bibitem[{{McLure} {et~al}\mbox{.}(2010){McLure}, {Dunlop}, {Cirasuolo},
  {Koekemoer}, {Sabbi}, {Stark}, {Targett}, \& {Ellis}}]{mclure2010}
{McLure} R.~J., {Dunlop} J.~S., {Cirasuolo} M., {Koekemoer} A.~M., {Sabbi} E.,
  {Stark} D.~P., {Targett} T.~A., {Ellis} R.~S., 2010, \mnras, 403, 960

\bibitem[{{McLure} {et~al}\mbox{.}(2011){McLure}, {Dunlop}, {de Ravel},
  {Cirasuolo}, {Ellis}, {Schenker}, {Robertson}, {Koekemoer}, {Stark}, \&
  {Bowler}}]{mclure2011}
{McLure} R.~J. {et~al.}, 2011, \mnras, 418, 2074

\bibitem[{{Moore} {et~al}\mbox{.}(1999{\natexlab{a}}){Moore}, {Ghigna},
  {Governato}, {Lake}, {Quinn}, {Stadel}, \& {Tozzi}}]{moore1999b}
{Moore} B., {Ghigna} S., {Governato} F., {Lake} G., {Quinn} T., {Stadel} J.,
  {Tozzi} P., 1999{\natexlab{a}}, \apjl, 524, L19

\bibitem[{{Moore} {et~al}\mbox{.}(1999{\natexlab{b}}){Moore}, {Quinn},
  {Governato}, {Stadel}, \& {Lake}}]{moore1999}
{Moore} B., {Quinn} T., {Governato} F., {Stadel} J., {Lake} G.,
  1999{\natexlab{b}}, \mnras, 310, 1147

\bibitem[{{Mori} {et~al}\mbox{.}(2002){Mori}, {Ferrara}, \& {Madau}}]{Mori2002}
{Mori} M., {Ferrara} A., {Madau} P., 2002, \apj, 571, 40

\bibitem[{{Nagamine} {et~al}\mbox{.}(2010){Nagamine}, {Ouchi}, {Springel}, \&
  {Hernquist}}]{nagamine2010}
{Nagamine} K., {Ouchi} M., {Springel} V., {Hernquist} L., 2010, \pasj, 62, 1455

\bibitem[{{Navarro} {et~al}\mbox{.}(1996){Navarro}, {Frenk}, \&
  {White}}]{navarro1996}
{Navarro} J.~F., {Frenk} C.~S., {White} S.~D.~M., 1996, \apj, 462, 563

\bibitem[{{Oesch} {et~al}\mbox{.}(2010{\natexlab{a}}){Oesch}, {Bouwens},
  {Carollo}, {Illingworth}, {Trenti}, {Stiavelli}, {Magee}, {Labb{\'e}}, \&
  {Franx}}]{oesch2010b}
{Oesch} P.~A. {et~al.}, 2010{\natexlab{a}}, \apjl, 709, L21

\bibitem[{{Oesch} {et~al}\mbox{.}(2010{\natexlab{b}}){Oesch}, {Bouwens},
  {Illingworth}, {Carollo}, {Franx}, {Labb{\'e}}, {Magee}, {Stiavelli},
  {Trenti}, \& {van Dokkum}}]{oesch2010}
{Oesch} P.~A. {et~al.}, 2010{\natexlab{b}}, \apjl, 709, L16

\bibitem[{{Oesch} {et~al}\mbox{.}(2013){Oesch}, {Bouwens}, {Illingworth},
  {Labb{\'e}}, {Franx}, {van Dokkum}, {Trenti}, {Stiavelli}, {Gonzalez}, \&
  {Magee}}]{oesch2013}
{Oesch} P.~A. {et~al.}, 2013, \apj, 773, 75

\bibitem[{{Ouchi} {et~al}\mbox{.}(2010){Ouchi}, {Shimasaku}, {Furusawa},
  {Saito}, {Yoshida}, {Akiyama}, {Ono}, {Yamada}, {Ota}, {Kashikawa}, {Iye},
  {Kodama}, {Okamura}, {Simpson}, \& {Yoshida}}]{ouchi2010}
{Ouchi} M. {et~al.}, 2010, \apj, 723, 869

\bibitem[{{Padovani} \& {Matteucci}(1993)}]{padovani-matteucci1993}
{Padovani} P., {Matteucci} F., 1993, \apj, 416, 26

\bibitem[{{Parkinson} {et~al}\mbox{.}(2008){Parkinson}, {Cole}, \&
  {Helly}}]{parkinson2008}
{Parkinson} H., {Cole} S., {Helly} J., 2008, \mnras, 383, 557

\bibitem[{{Peng} {et~al}\mbox{.}(2010){Peng}, {Lilly}, {Kova{\v c}},
  {Bolzonella}, {Pozzetti}, {Renzini}, {Zamorani}, {Ilbert}, {Knobel},
  {Iovino}, {Maier}, {Cucciati}, {Tasca}, {Carollo}, {Silverman}, {Kampczyk},
  {de Ravel}, {Sanders}, {Scoville}, {Contini}, {Mainieri}, {Scodeggio},
  {Kneib}, {Le F{\`e}vre}, {Bardelli}, {Bongiorno}, {Caputi}, {Coppa}, {de la
  Torre}, {Franzetti}, {Garilli}, {Lamareille}, {Le Borgne}, {Le Brun},
  {Mignoli}, {Perez Montero}, {Pello}, {Ricciardelli}, {Tanaka}, {Tresse},
  {Vergani}, {Welikala}, {Zucca}, {Oesch}, {Abbas}, {Barnes}, {Bordoloi},
  {Bottini}, {Cappi}, {Cassata}, {Cimatti}, {Fumana}, {Hasinger}, {Koekemoer},
  {Leauthaud}, {Maccagni}, {Marinoni}, {McCracken}, {Memeo}, {Meneux}, {Nair},
  {Porciani}, {Presotto}, \& {Scaramella}}]{peng2010}
{Peng} Y.-j. {et~al.}, 2010, \apj, 721, 193

\bibitem[{{Planck Collaboration}(2013)}]{planck20132}
{Planck Collaboration} ., 2013, ArXiv e-prints:1303.5076

\bibitem[{{Planck Collaboration} {et~al}\mbox{.}(2013){Planck Collaboration},
  {Ade}, {Aghanim}, {Armitage-Caplan}, {Arnaud}, {Ashdown}, {Atrio-Barandela},
  {Aumont}, {Baccigalupi}, {Banday}, \& et~al.}]{planck2013}
{Planck Collaboration} {et~al.}, 2013, ArXiv e-prints:1303.5062

\bibitem[{{Postman} {et~al}\mbox{.}(2012){Postman}, {Coe}, {Ben{\'{\i}}tez},
  {Bradley}, {Broadhurst}, {Donahue}, {Ford}, {Graur}, {Graves}, {Jouvel},
  {Koekemoer}, {Lemze}, {Medezinski}, {Molino}, {Moustakas}, {Ogaz}, {Riess},
  {Rodney}, {Rosati}, {Umetsu}, {Zheng}, {Zitrin}, {Bartelmann}, {Bouwens},
  {Czakon}, {Golwala}, {Host}, {Infante}, {Jha}, {Jimenez-Teja}, {Kelson},
  {Lahav}, {Lazkoz}, {Maoz}, {McCully}, {Melchior}, {Meneghetti}, {Merten},
  {Moustakas}, {Nonino}, {Patel}, {Reg{\"o}s}, {Sayers}, {Seitz}, \& {Van der
  Wel}}]{postman2012}
{Postman} M. {et~al.}, 2012, \apjs, 199, 25

\bibitem[{{Rogers} {et~al}\mbox{.}(2014){Rogers}, {McLure}, {Dunlop}, {Bowler},
  {Curtis-Lake}, {Dayal}, {Faber}, {Ferguson}, {Finkelstein}, {Grogin},
  {Hathi}, {Kocevski}, {Koekemoer}, \& {Kurczynski}}]{rogers2014}
{Rogers} A.~B. {et~al.}, 2014, \mnras, 440, 3714

\bibitem[{{Salvaterra} {et~al}\mbox{.}(2011){Salvaterra}, {Ferrara}, \&
  {Dayal}}]{salvaterra2011}
{Salvaterra} R., {Ferrara} A., {Dayal} P., 2011, \mnras, 414, 847

\bibitem[{{Sheth} \& {Tormen}(1999)}]{sheth-tormen1999}
{Sheth} R.~K., {Tormen} G., 1999, \mnras, 308, 119

\bibitem[{{Smit} {et~al}\mbox{.}(2014){Smit}, {Bouwens}, {Labb{\'e}}, {Zheng},
  {Bradley}, {Donahue}, {Lemze}, {Moustakas}, {Umetsu}, {Zitrin}, {Coe},
  {Postman}, {Gonzalez}, {Bartelmann}, {Ben{\'{\i}}tez}, {Broadhurst}, {Ford},
  {Grillo}, {Infante}, {Jimenez-Teja}, {Jouvel}, {Kelson}, {Lahav}, {Maoz},
  {Medezinski}, {Melchior}, {Meneghetti}, {Merten}, {Molino}, {Moustakas},
  {Nonino}, {Rosati}, \& {Seitz}}]{smit2014}
{Smit} R. {et~al.}, 2014, \apj, 784, 58

\bibitem[{{Somerville} \& {Primack}(1999)}]{somerville-primack1999}
{Somerville} R.~S., {Primack} J.~R., 1999, \mnras, 310, 1087

\bibitem[{{Springel} \& {Hernquist}(2003)}]{springel-hernquist2003b}
{Springel} V., {Hernquist} L., 2003, \mnras, 339, 289

\bibitem[{{Springel} {et~al}\mbox{.}(2008){Springel}, {Wang}, {Vogelsberger},
  {Ludlow}, {Jenkins}, {Helmi}, {Navarro}, {Frenk}, \& {White}}]{springel2008}
{Springel} V. {et~al.}, 2008, \mnras, 391, 1685

\bibitem[{{Stark} {et~al}\mbox{.}(2009){Stark}, {Ellis}, {Bunker}, {Bundy},
  {Targett}, {Benson}, \& {Lacy}}]{stark2009}
{Stark} D.~P., {Ellis} R.~S., {Bunker} A., {Bundy} K., {Targett} T., {Benson}
  A., {Lacy} M., 2009, \apj, 697, 1493

\bibitem[{{Stark} {et~al}\mbox{.}(2013){Stark}, {Schenker}, {Ellis},
  {Robertson}, {McLure}, \& {Dunlop}}]{stark2013}
{Stark} D.~P., {Schenker} M.~A., {Ellis} R., {Robertson} B., {McLure} R.,
  {Dunlop} J., 2013, \apj, 763, 129

\bibitem[{{Stinson} {et~al}\mbox{.}(2013){Stinson}, {Brook}, {Macci{\`o}},
  {Wadsley}, {Quinn}, \& {Couchman}}]{stinson2013}
{Stinson} G.~S., {Brook} C., {Macci{\`o}} A.~V., {Wadsley} J., {Quinn} T.~R.,
  {Couchman} H.~M.~P., 2013, \mnras, 428, 129

\bibitem[{{Tornatore} {et~al}\mbox{.}(2007){Tornatore}, {Borgani}, {Dolag}, \&
  {Matteucci}}]{tornatore2007}
{Tornatore} L., {Borgani} S., {Dolag} K., {Matteucci} F., 2007, \mnras, 382,
  1050

\bibitem[{{White} \& {Frenk}(1991)}]{white-frenk1991}
{White} S.~D.~M., {Frenk} C.~S., 1991, \apj, 379, 52

\bibitem[{{Wilkins} {et~al}\mbox{.}(2011){Wilkins}, {Bunker}, {Stanway},
  {Lorenzoni}, \& {Caruana}}]{wilkins2011}
{Wilkins} S.~M., {Bunker} A.~J., {Stanway} E., {Lorenzoni} S., {Caruana} J.,
  2011, \mnras, 417, 717

\bibitem[{{Yoshida} {et~al}\mbox{.}(2002){Yoshida}, {Stoehr}, {Springel}, \&
  {White}}]{yoshida2002}
{Yoshida} N., {Stoehr} F., {Springel} V., {White} S.~D.~M., 2002, \mnras, 335,
  762

\bibitem[{{Zheng} {et~al}\mbox{.}(2012){Zheng}, {Postman}, {Zitrin},
  {Moustakas}, {Shu}, {Jouvel}, {H{\o}st}, {Molino}, {Bradley}, {Coe},
  {Moustakas}, {Carrasco}, {Ford}, {Ben{\'{\i}}tez}, {Lauer}, {Seitz},
  {Bouwens}, {Koekemoer}, {Medezinski}, {Bartelmann}, {Broadhurst}, {Donahue},
  {Grillo}, {Infante}, {Jha}, {Kelson}, {Lahav}, {Lemze}, {Melchior},
  {Meneghetti}, {Merten}, {Nonino}, {Ogaz}, {Rosati}, {Umetsu}, \& {van der
  Wel}}]{zheng2012}
{Zheng} W. {et~al.}, 2012, \nat, 489, 406

\end{thebibliography}

\label{lastpage} 
\end{document}